\documentclass[showpacs, prb,twocolumn,preprintnumbers , superscriptaddress, aps]{revtex4-2}

\usepackage{color}
\usepackage{amsmath,amssymb}
\usepackage{pifont}
\usepackage{amssymb}  
\usepackage{bbold}
\usepackage{float}
\usepackage{stfloats}
\usepackage[normalem]{ulem}

\usepackage[labelfont=bf]{caption} 
\usepackage{tikz}
\usepackage{bbold}
\usepackage{makecell}
\usepackage{pifont}   
\usepackage{graphicx} 
\usepackage{dcolumn}  
\usepackage{bm}       
\usepackage{multirow} 
\usepackage{placeins}
\usepackage[colorlinks]{hyperref}
\usepackage{mathtools}
\usepackage{subcaption}
\usepackage{amsmath}  

\usepackage[normalem]{ulem}

\usepackage{mathrsfs}

\usepackage{multirow}
\usepackage{float}

\begin{document}
\title{Local Pressure-Induced Quantum Dot-to-Ring Transition and Aharonov–Bohm Effect in Reconfigurable $\rm MoTe_2$ Bilayers}
\author{Ibsal Assi}
\email{iassi@mun.ca}
\affiliation{Department of Physics and Physical Oceanography,
Memorial University of Newfoundland and Labrador, St. John’s, Newfoundland \& Labrador, Canada A1B 3X7}

\author{A. Belayadi}
\affiliation{Department of theoretical physics$,$ University of Science and Technology Houari Boumediene$,$ Bab-Ezzouar, Algeria.}

\author{Hocine Bahlouli}
\affiliation{Physics Department and Interdisciplinary Research Center for Advanced Materials,
		King Fahd University of Petroleum \& Minerals, Dhahran 31261, Saudi Arabia}

\author{Udo Schwingenschl\"ogl}
\email{udo.schwingenschlogl@kaust.edu.sa}
\affiliation{Physical Science and Engineering Division, King Abdullah University of Science and Technology, Thuwal 23955-6900, Saudi Arabia}

\author{J. P. F. LeBlanc}
\affiliation{Department of Physics and Physical Oceanography,
Memorial University of Newfoundland and Labrador, St. John’s, Newfoundland \& Labrador, Canada A1B 3X7}
\affiliation{Compute Everything Technologies Ltd., St. John's, Newfoundland \& Labrador, Canada}
\date{\today}
\begin{abstract}
We demonstrate that local pressure provides an effective means of engineering reconfigurable quantum confinement in MoTe$_2$ homobilayers. By applying a smooth, spatially varying pressure profile, we show that the confinement geometry can be continuously tuned from a conventional quantum dot to a quantum ring, with the emergence of annular confinement being a distinctive feature of AB-stacked bilayers. This crossover originates from the pressure-induced spatial modulation in the electronic parameters, which produces an off-center minimum in the effective confinement potential, thereby redistributing the ground-state probability density away from the center and giving rise to an annular confinement geometry. In the quantum-ring regime, a perpendicular magnetic field gives rise to a flux-periodic energy spectrum and persistent currents, providing characteristic signatures of the Aharonov–Bohm effect. By coupling the pressure-induced quantum ring to source and drain leads, we further demonstrate pronounced Aharonov–Bohm conductance oscillations, accompanied by real-space current patterns exhibiting constructive and destructive interference at integer and half-integer magnetic flux quanta, respectively. These findings establish locally pressurized MoTe$_2$ bilayers as a lithography-free platform for reconfigurable quantum confinement and realizing pressure-tunable Aharonov-Bohm interferometry in two dimensional materials.
\end{abstract}

\maketitle

\section{Introduction}

Quantum confinement in two-dimensional (2D) materials has attracted considerable attention due to its potential applications in quantum information processing, nano-electronics, and spin and valley-based devices \cite{Schaibley2016-xq,Kim2024-nw,Kim2024-gz}. In particular, semiconductor quantum dots (QDs) provide a flexible platform for realizing discrete electronic states whose energies and spatial localization can be controlled by external electric, magnetic, and mechanical perturbations \cite{Kouwenhoven2001-di,RevModPhys.79.1217,PhysRevLett.105.067404}. The emergence of van der Waals materials and their layered structures has substantially expanded the possibilities for engineering quantum confinement at the atomic scale \cite{Liu2016-ea,Novoselov2016-yl,Geim2013-nz}. Their reduced dimensionality and sensitivity to stacking, strain, and external fields provide a highly controllable setting in which localized electronic states can be created and manipulated \cite{Geim2013-nz,PhysRevB.97.195454,Chirolli2019-ze}.

Among these materials, transition-metal dichalcogenide (TMD) bilayers have attracted considerable attention as promising systems for engineering and controlling electronic states \cite{Xu2014-aa,PhysRevB.110.115410}. Their substantial band gaps, strong spin-orbit coupling, and significant interlayer hybridization give rise to a rich variety of electronic phenomena that are strongly affected by stacking configuration, twist angle, and external perturbations \cite{Liu2015-md,PhysRevB.89.075409,PhysRevB.102.235418}. In twisted TMD homobilayers, the resulting moiré superlattice offers an additional mean for controlling the low-energy electronic structure, enabling the emergence of flat bands and nontrivial topological phases \cite{PhysRevLett.122.086402,PhysRevB.105.L081108,PhysRevLett.132.096602}. Pressure provides yet another powerful control parameter by altering the interlayer separation and, consequently, the effective masses, interlayer tunneling amplitudes, and moiré potential parameters \cite{miftah,PhysRevB.99.024111,th65-gff6}. In twisted MoTe$_2$ bilayers, theoretical studies have shown that such pressure-induced modifications can substantially reshape the low-energy bands and drive transitions between distinct topological regimes, including changes in valley Chern numbers and the formation of extremely flat moiré bands \cite{miftah,4742-rldh}.

Although the influence of pressure on the electronic structure and moiré-band topology of twisted MoTe$_2$ has recently attracted growing attention, its role in producing quantum confinement remains comparatively unexplored \cite{miftah,4742-rldh}. Local strain has been successfully used to generate localized states and quantum emitters in two-dimensional semiconductors. However, to our knowledge, the use of spatially varying pressure to create and manipulate single-particle quantum-dots through pressure-dependent electronic parameters has received little attention \cite{Kim2026-sq,Branny2017-nz}. Investigating such confined states is important not only for understanding their fundamental physical properties but also for developing quantum devices, in which localized charge carriers serve as essential building blocks of quantum dot architectures \cite{loss1998,RevModPhys.79.1217,RevModPhys.95.025003}.

Several approaches have been developed to generate QDs in 2D materials, including electrostatic gating, defect engineering, and strain-induced confinement \cite{PhysRevApplied.13.054058,M_C_de_Melo2021-kq,Dirnberger2021-uc}. In TMD bilayers, local mechanical deformation provides an additional means of engineering quantum confinement by simultaneously modifying the local electronic structure and interlayer coupling \cite{Kim2026-sq,Koo2024-qu,Maiti2021-ps}. Motivated by recent advances in nanoscale pressure probes and local strain engineering, we investigate whether local pressure can be exploited to generate and continuously tune confined states in MoTe$_2$ homobilayers \cite{Ai2025-fe}.

To address this question, we study quantum confinement in MoTe$_2$ homobilayers using a pressure-dependent continuum model recently derived from first-principles calculations \cite{miftah}. For comparison, we consider two confinement mechanisms: a Gaussian confinement potential representing conventional electrostatic quantum dot (QD) confinement, and a spatially varying pressure profile that generates confinement through local modulation of pressure-dependent electronic parameters. Unlike purely electrostatic confinement, the pressure-induced mechanism modifies the underlying electronic structure, providing an additional means to control both the spatial extent and geometry of the confined states. We show that this mechanism produces a stacking-selective crossover from quantum-dot to quantum-ring confinement: annular localization emerges in the AB-stacked bilayer as the pressure amplitude increases, whereas the AA-stacked system remains predominantly quantum-dot-like. In the ring regime, a perpendicular magnetic field produces Aharonov-Bohm oscillations in the energy spectrum and persistent currents. By coupling the pressure-defined ring to source and drain leads, we further observe flux-periodic conductance oscillations accompanied by real-space current patterns exhibiting constructive and destructive quantum interference, thereby establishing the system as a pressure-tunable Aharonov-Bohm interferometer.

The remainder of this paper is organized as follows. In Sec. II, we introduce the pressure-dependent continuum model and describe the confinement mechanisms considered in this work. In Sec. III, we present our numerical results, focusing on the pressure-induced QD to quantum ring transition in AB-stacked MoTe$_2$, its Aharonov-Bohm response, the corresponding behavior of the AA-stacked bilayer, and comparison between pressure-induced and static confinement. We then investigate quantum transport through a pressure-defined ring coupled to source and drain leads, demonstrating its functionality as a tunable Aharonov–Bohm interferometer. Finally, Sec. IV summarizes our main findings, while additional numerical details are provided in the Appendix.

\section{Mathematical Model}

\subsection{Continuum Low Energy Hamiltonian}
We begin by introducing the continuum low-energy model for electrons in a MoTe$_2$ homobilayer. Each monolayer consists of a two-dimensional hexagonal lattice, whose Brillouin zone contains two inequivalent $K$ and $K'$ valleys. Focusing on the $K$ valley, we adopt the pressure-dependent continuum Hamiltonian recently derived from density functional theory calculations in Ref. \cite{miftah}, 
\begin{equation}
\label{eq:MainH}
    H = \begin{pmatrix}
 -\frac{\hbar^2}{2m^*} k^2 + \Delta_{t}& \Delta_T \\
\Delta_T^\dagger &   -\frac{\hbar^2}{2m^*} k^2 + \Delta_{b}
\end{pmatrix}
\end{equation}
where $\hbar$ is the reduced Planck's constant, $m^*$ is the electron's effective mass, $k$ is the electron's crystal momentum, $\Delta_{t/b}$ is the intra-layer hopping, and $\Delta_T$ is the inter-layer hopping. The parameters appearing in our Hamiltonian Eq. \ref{eq:MainH} include functions of the applied pressure which were obtained through fitting \cite{miftah}, details of which will be discussed below. First, the effective mass varies with pressure according to \cite{miftah}
\begin{equation}\label{eq_2}
    m^*(p) = (0.640 - 0.193 p + 0.0335 p^2) m_e
\end{equation}
where $m_e$ is the free electron mass and $p$ is the pressure in units of GPa. The intra-layer hopping $\Delta_{t/b}$ is given by 
\begin{equation}
\label{eqn:Delta_ell}
\Delta_\ell= \sum_{j=1,3,5} 2V(p) \cos(\vec{G}_j \cdot \vec{d} + \ell \phi).
\end{equation}
with $\ell=-1$ ($+1$) for the top (bottom) layer and $\vec{d}$ is a displacement vector. The pressure dependent intra-layer coupling strength is set by parameter $V(p)$ that has the form \cite{miftah}
\begin{equation}
\label{eq:Vp}
    V(p) = \frac{a + b p + c p^2 + d p^3 + e p^4 + f p^5}{g + h p + i p^2 + j p^3 + k p^4 + l p^5}
\end{equation}
with the fitted coefficients taken directly from Ref. \cite{miftah},
\begin{equation}
\begin{aligned}
a &= -694, \quad b = 258, \quad c = 1327, \quad d = -1405, \\
e &= 486.1, \quad f = -55.8, \quad g = -80.5, \quad h = 79.7, \\
i &= 61.9, \quad j = -102.3, \quad k = 39.51, \quad l = -4.783
\end{aligned}
\end{equation}
and $\vec{G}_j = R\left(\frac{(j-1)\pi}{3}\right) \frac{4\pi}{a_0 \sqrt{3}} \hat{j}
$ is the reciprocal lattice vector with the phase $\phi=-90.0797^\circ$. Finally, the inter-layer hopping is given as \cite{miftah}
\begin{equation}
\label{eqn:Delta_T}
\Delta_T=w(p) \left( 1 + e^{-i \vec{G}_2 \cdot \vec{d}} + e^{-i \vec{G}_3 \cdot \vec{d}} \right)
\end{equation}
where $w(p) = (-8.35 - 1.58p + 0.25p^2) \, \text{meV}$ is the pressure dependent scale that determines the strength of the inter-layer hopping. We emphasize that the model parameters provided are valid for a pressure range between zero and $3.5$ GPa. Finally, the displacement vector $\vec{d}$ that interpolates between different high symmetry stacking configurations is defined by
\begin{equation}
    \vec{d} = \frac{n(\vec{a}_2-\vec{a}_1)}{3}, \quad n = 0, \pm 1.
\end{equation}
where
\begin{equation}
    \vec{a}_1=a_0(1,0),\quad \vec{a}_2=\frac{a_0}{2}(-1,\sqrt{3})
\end{equation}
are the primitive lattice vectors with $a_0=0.3487$ nm. Here $n=0$ corresponds to so-called AA stacking, where atoms from one layer are exactly on top of similar atoms on the other layer and $n=+1(-1)$ corresponds to so-called AB(BA) stacking, where atoms from one layer lie exactly above the empty hexagon sites on the other layer. 

In the next section, we will use this model to investigate quantum confinement induced by both local pressure and static confining potentials.

\subsection{Introducing confinement}
We introduce two ways to confine electrons within the $\mathbf{\mathrm{MoTe}_2}$ homobilayer. First, using local pressure modeled by a Gaussian profile,
\begin{equation}
    p(r)=p_0 e^{-\frac{r^2}{\alpha^2_p}}
\end{equation}
where $p_0$ is the pressure amplitude, taking values within the range 0-3.5 GPa \cite{miftah}, $r$ is the radial distance from the center of the region where the probe acts on the material, and $\alpha_p$ is the characteristic length scale of the pressure distribution. This local pressure model modifies the interlayer and intralayer hopping, as well as the effective mass making them position dependent. However, the latter requires a more careful treatment since $\frac{\hbar^2k^2}{2m^*}$ is not Hermitian when $m^*$ is position-dependent. To preserve Hermiticity of the kinetic energy operator, we apply a symmetrized canonical replacement $\frac{\hbar^2k^2}{2m^*}\to \frac{\hbar^2}{2} \vec{k}\cdot \left( \frac{1}{m^*}\right) \vec{k}$ with $\vec{k}=-i\vec{\nabla}$. 

An alternative way to confine electrons is by introducing a static potential. For simplicity, we consider a Gaussian confining potential of the form
\begin{equation}
\label{eqn:confinement}
U(r)=U_0e^{-\frac{r^2}{\alpha^2_U}}
\end{equation}
where $U_0>0$ is the potential depth and $\alpha_U$ the characteristic width of the confining potential. We assume that the electrostatic confinement and local pressure profile possess the same characteristic spatial extent, $\alpha_p=\alpha_U=\alpha$. Since the two perturbations originate from independent physical mechanisms, relaxing this assumption would quantitatively modify the confinement but is not expected to alter the qualitative physics.

Thus, the Hamiltonian describing the $K$-valley electrons in the $\mathbf{\mathrm{MoTe}_2}$ bilayer under the influence of both local pressure and static confining potential is
\begin{equation}
\label{eq:H_QD}
    H_{\rm QD}(r)=\begin{pmatrix}
 -\hat{T}+ \Delta_{t}(r)+U(r)& \Delta_T(r)\\
\Delta_T^\dagger(r)&   -\hat{T}  + \Delta_{b}(r)+U(r)
\end{pmatrix}
\end{equation}
where $\hat{T}=\frac{\hbar^2}{2} \vec{k}\cdot \left( \frac{1}{m^*(r)} \vec{k}\right)$ and $\Delta_{t,b,T}(r)$ are obtained from Eqs. \eqref{eqn:Delta_ell} and \eqref{eqn:Delta_T} using the position dependent pressure. The procedure for diagonalizing this Hamiltonian is described in Appendix \ref{sec:num_diag}. 

\begin{figure}[h!]
    \centering
    \includegraphics[width=\linewidth]{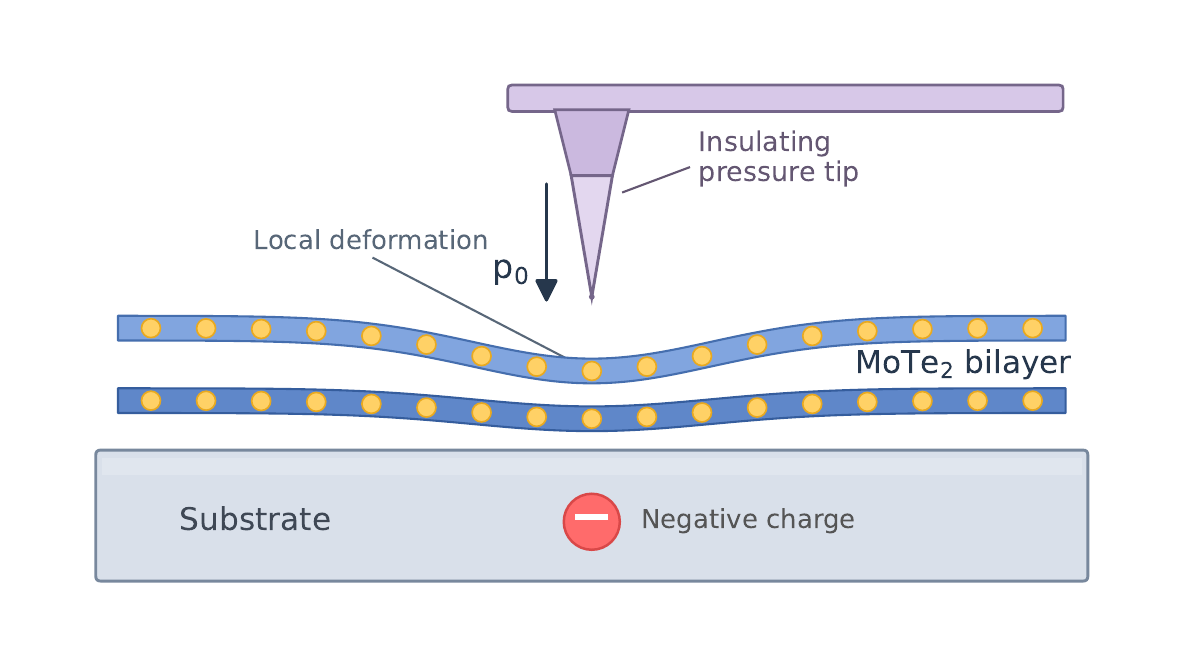}
    \caption{Schematic of the modeled geometry. The ${\rm MoTe_2}$ bilayer placed on top of a dielectric substrate carrying a negative charge and is under the influence of pressure tip.}
    \label{fig:device}
\end{figure}

To realize these effects, the ${\rm MoTe_2}$ bilayer is vertically stacked on a dielectric substrate hosting a localized negative charge distribution, which generates a static confining potential for electrons. Local mechanical pressure is then applied to the top layer using a pressure probe, such as an electrically insulating tip. This combined arrangement of electrostatic confinement from the substrate and local pressure from the tip is illustrated schematically in Fig.~\ref{fig:device}. The electron confinement can be tuned by varying the pressure amplitude, tip size, and the substrate charge.

\section{Results and Discussions}

\subsection{Quantum dot-to-ring crossover in AB-stacked ${\rm MoTe_2}$}
\label{sec:dot_to_ring}
\begin{figure}[h!]
    \centering
    \includegraphics[width=\linewidth]{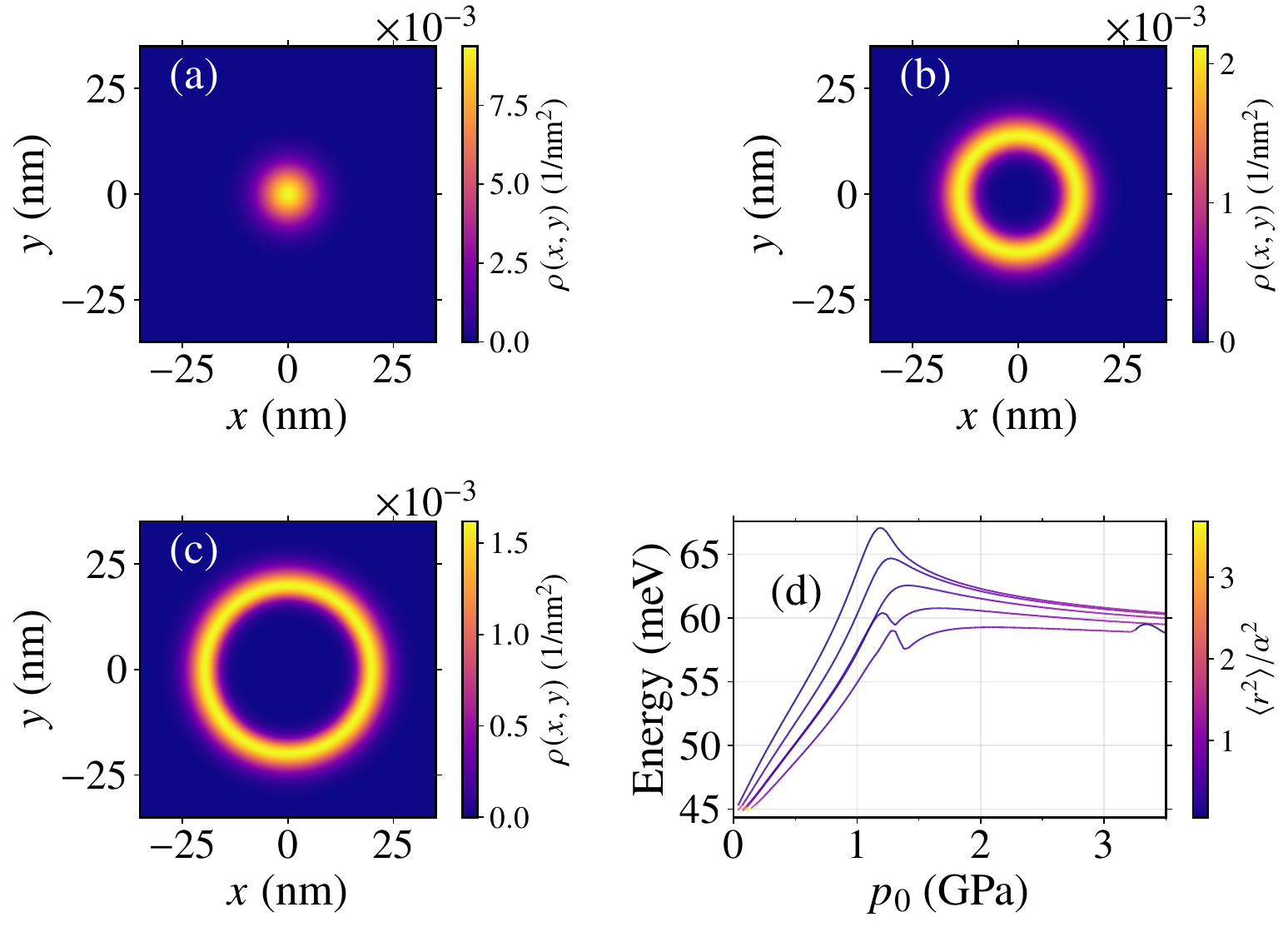}
    \caption{GS density $\rho(x,y)$ of the AB-stacked ${\rm MoTe_2}$ bilayer at local pressure amplitudes (a) $p_0=1.0$ GPa, (b) $p_0=1.5$ GPa and (c) $p_0=2.0$ GPa. (d) Five distinct localized energy levels as functions of $p_0$; the highest-energy branch is the nondegenerate GS, while each of the other four branches is twofold degenerate. Parameters: $\alpha=25$ GPa and $U_0=0$.}
    \label{fig:dot_to_ring_AB_stacked}
\end{figure}

First, we will investigate the local pressure confinement in the ${\rm MoTe_2}$ bilayer without static confinement ($U_0=0$). In doing so, we consider the ground state at $p_0=1$ GPa, $p_0=1.5$ GPa, and $p_0=2$ GPa as shown in Fig. \ref{fig:dot_to_ring_AB_stacked}(a), (b), and (c), respectively. At low pressure, the ground state (GS) is localized at the center ($r=0$), as the pressure increases, the GS is repelled away from the center and gets confined in annular region as shown in the probability density plots \ref{fig:dot_to_ring_AB_stacked}(b,c). In what follows, we use the term \textit{quantum ring} to denote such an annular electronic confinement state, in which the two-dimensional probability density becomes depleted at the center and develops a maximum at a finite radius, $r_{\rm peak}>0$, signaling the emergence of a ring-like confined state. The underlying MoTe$_2$ bilayer remains continuous; therefore, the ring refers to the geometry of the confined electronic state rather than to a physically patterned annulus. The crossover from quantum-dot to quantum-ring confinement occurs at a pressure of approximately 1.3 GPa.

\begin{figure}[h!]
    \centering
    \includegraphics[width=\linewidth]{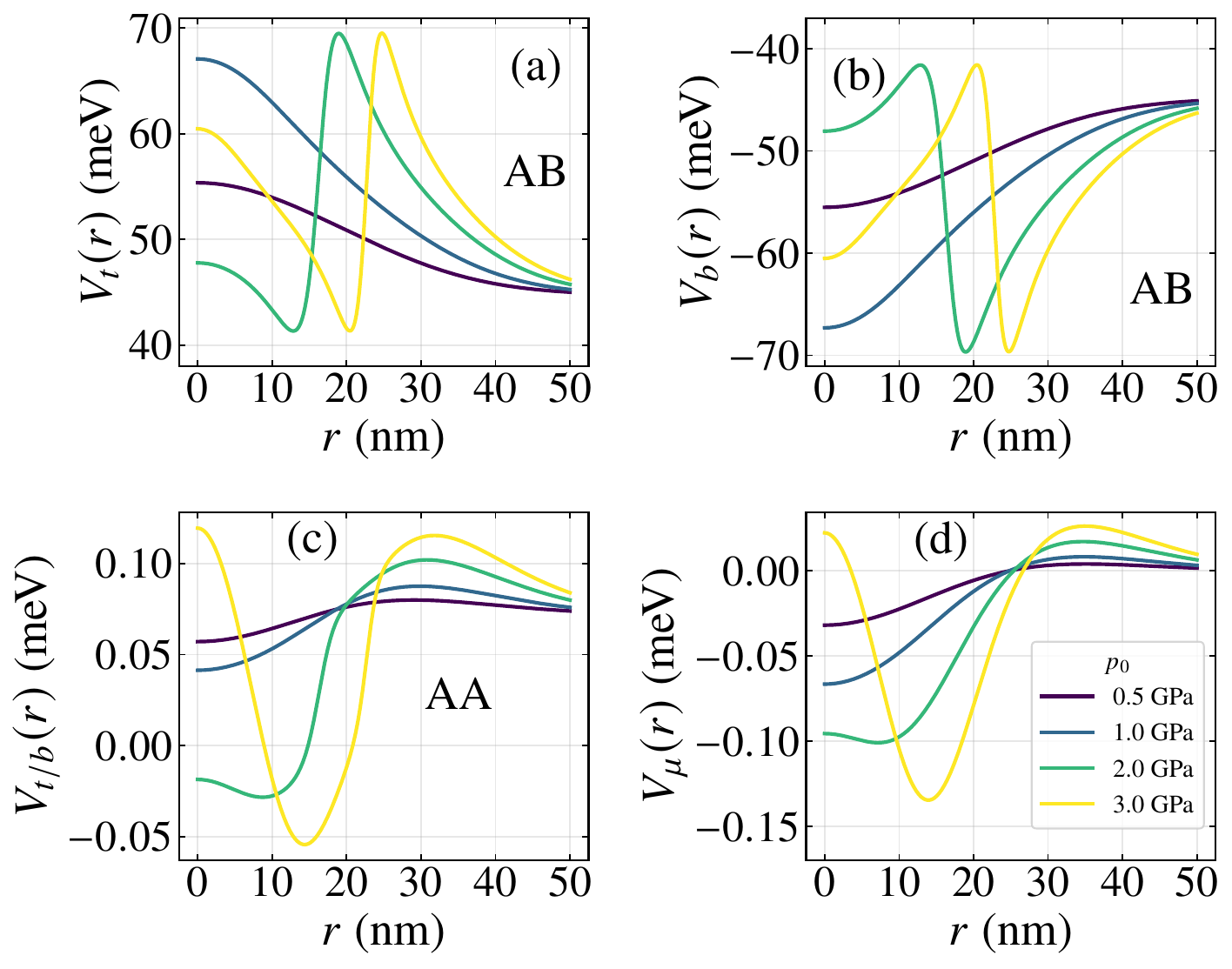}
    \caption{A plot of the effective potentials (a) $V_t(r)$ and (b) $V_b(r)$ [Eq. \eqref{eqn:V_tb}] for the AB-stacked $\mathrm{MoTe}_2$ bilayer. (c) The effective potentials $V_{t/b}(r)$ for the AA stacking configuration. (d) The geometric potential $V_\mu(r)$ [Eq. \eqref{eqn:V_mu}] produced by the spatial variations of the effective mass. Parameters: $\alpha=25$ nm.}
    \label{fig:effective_potentials_AB}
\end{figure}

To understand the origin of this crossover, we plot the effective potential [Eq. \eqref{eqn:V_tb}] in Fig. \ref{fig:effective_potentials_AB}. We notice that the effective potentials at larger pressure values develop two extrema with strong off-center minima that compensate for the electron's kinetic energy, which explains why the GS moves away from the center. Consequently, the ground-state probability density is progressively depleted at the origin and becomes concentrated around the off-center minima, driving the crossover from quantum-dot to quantum-ring confinement.This transition is solely due to the local pressure modifying the intra-layer hopping as shown in Fig. \ref{fig:effective_potentials_AB} (a,b) in which the kinetic energy scale is dominant meanwhile the geometric potential [Eq. \eqref{eqn:V_mu}] emerging from the spatial variations of the effective mass produces small perturbations as shown in Fig. \ref{fig:effective_potentials_AB}(d).

For further insight into this crossover, we consider the ground state (GS) and the first four distinct excited-state energy levels and plot their energies as functions of  $p_0$ in Fig. \ref{fig:dot_to_ring_AB_stacked}(d). The color scale indicates the expectation value of $r^2$, normalized by $\alpha^2$, $\langle r^2\rangle/\alpha^2$, and thus provides a quantitative measure of the spatial extent of each state. Notably, for the GS, $\langle r^2\rangle/\alpha^2$ exhibits a pronounced increase for $p_0 \gtrsim 1.3$ GPa, signaling an outward redistribution of the probability density. This behavior coincides with the onset of annular confinement observed in the density profiles of Figs.\ref{fig:dot_to_ring_AB_stacked}(b) and \ref{fig:dot_to_ring_AB_stacked}(c), providing further evidence for the pressure-driven crossover from quantum-dot to quantum-ring confinement.

\begin{figure}[h!]
    \centering
    \includegraphics[width=\linewidth]{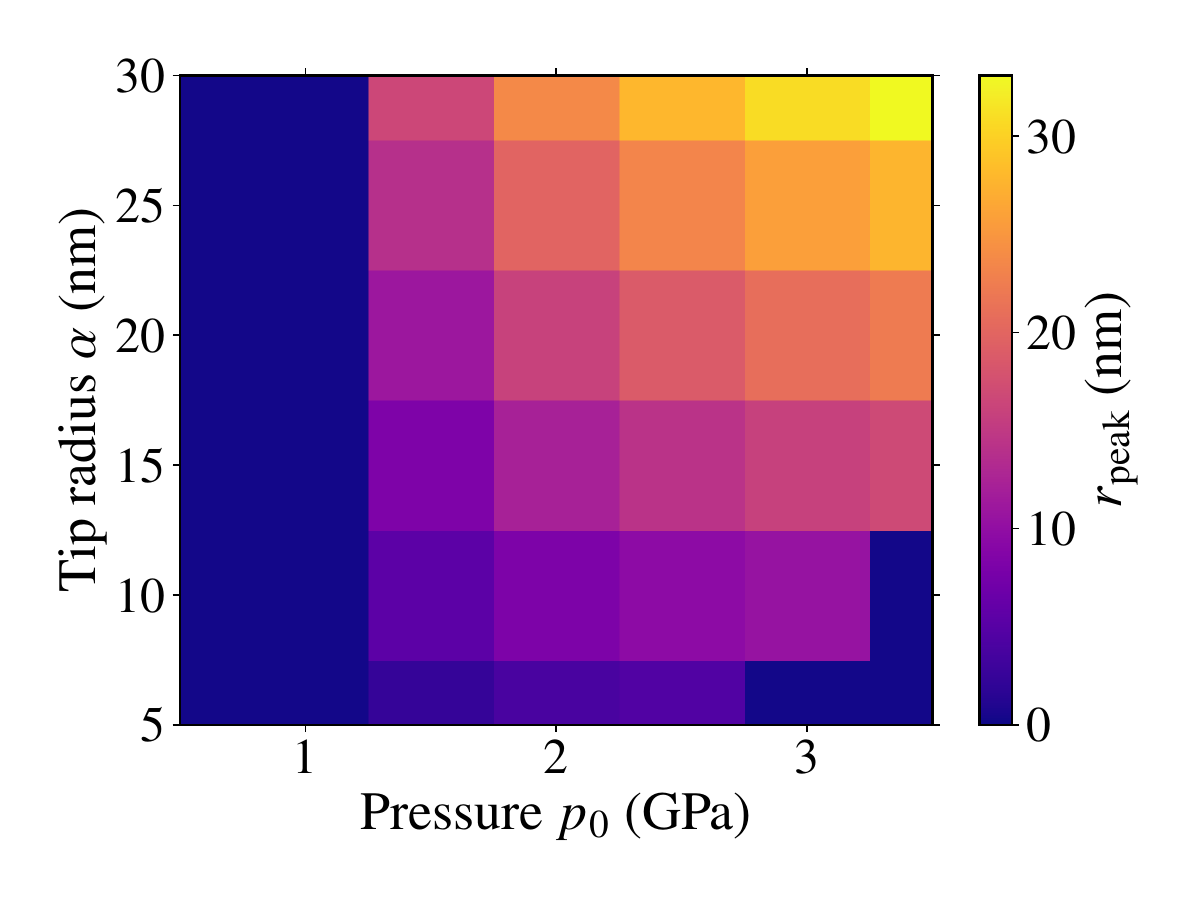}
    \caption{Location of the maximum of the probability density $\rho(x,y)=\left|\Psi(r,\theta)\right|^2$ as a function of $p_0$ and $\alpha$ for the AB-stacked $\mathrm{MoTe}_2$ bilayer.}
    \label{fig:phase_rpeak}
\end{figure}

To better elucidate the formation of the annular confinement, we plotted the center of the maxima in the probability density $\rho(x,y)=\left|\Psi(r,\theta)\right|^2$, denoted by $r_{\rm peak}$, as a function of $p_0$ and $\alpha$ in Fig. \ref{fig:phase_rpeak}. The resulting phase diagram clearly reveals that annular confinement becomes increasingly pronounced with increasing pressure and larger values of $\alpha$, as evidenced by the progressive shift of the probability-density maximum away from the center.

\begin{figure}
    \centering
    \includegraphics[width=\linewidth]{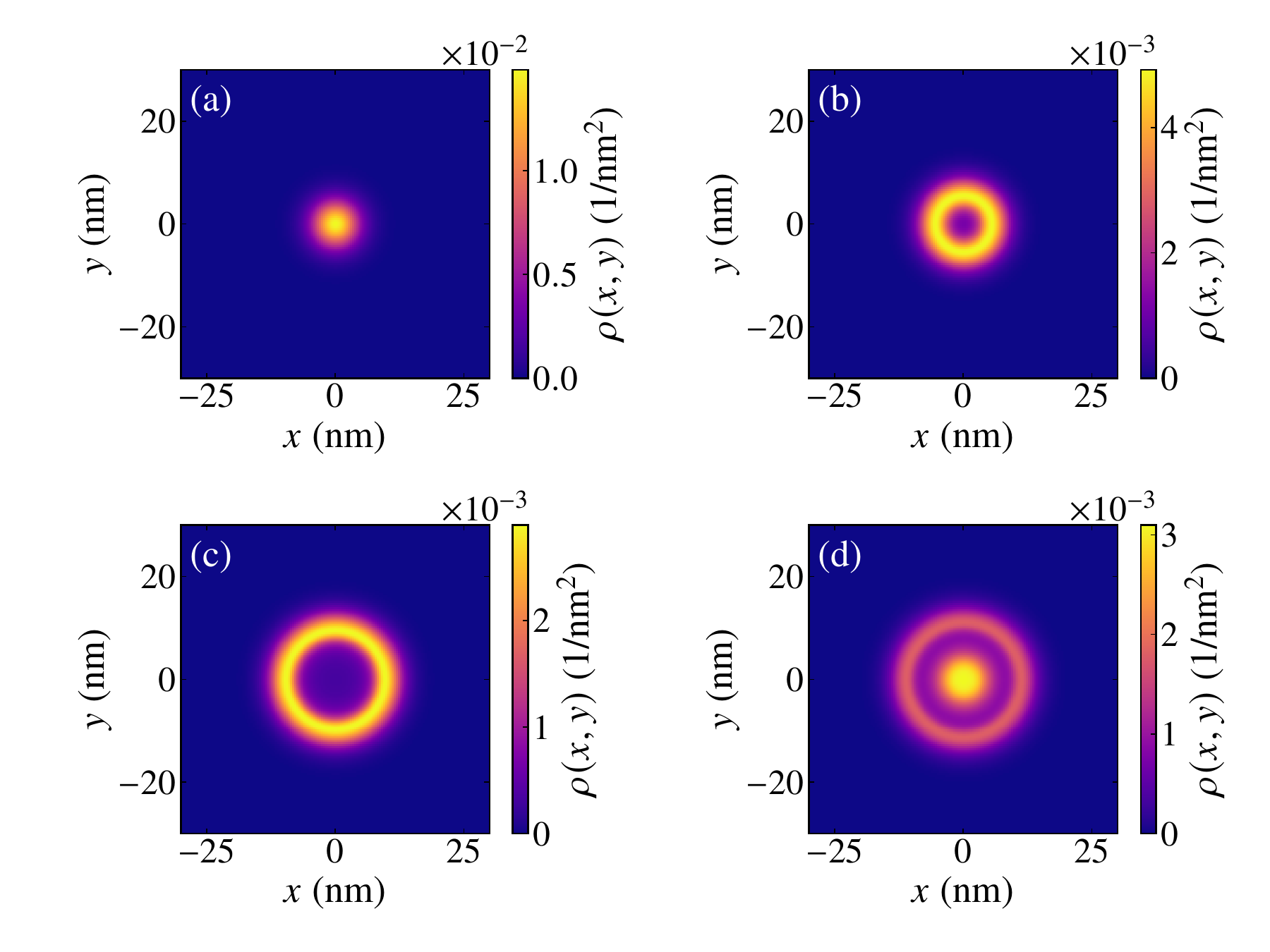}
    \caption{Plot of the probability density of the AB-stacked $\mathrm{MoTe}_2$ bilayer for the pressure values (a) $p_0=1.0$ GPa, (b) $p_0=1.5$ GPa, (c) $p_0=2.5$ GPa, and (d)$p_0=3.5$ GPa. Parameters:  $\alpha=10$ nm. }
    \label{fig:AB_alpha10nm}
\end{figure}

\begin{figure}
    \centering
    \includegraphics[width=\linewidth]{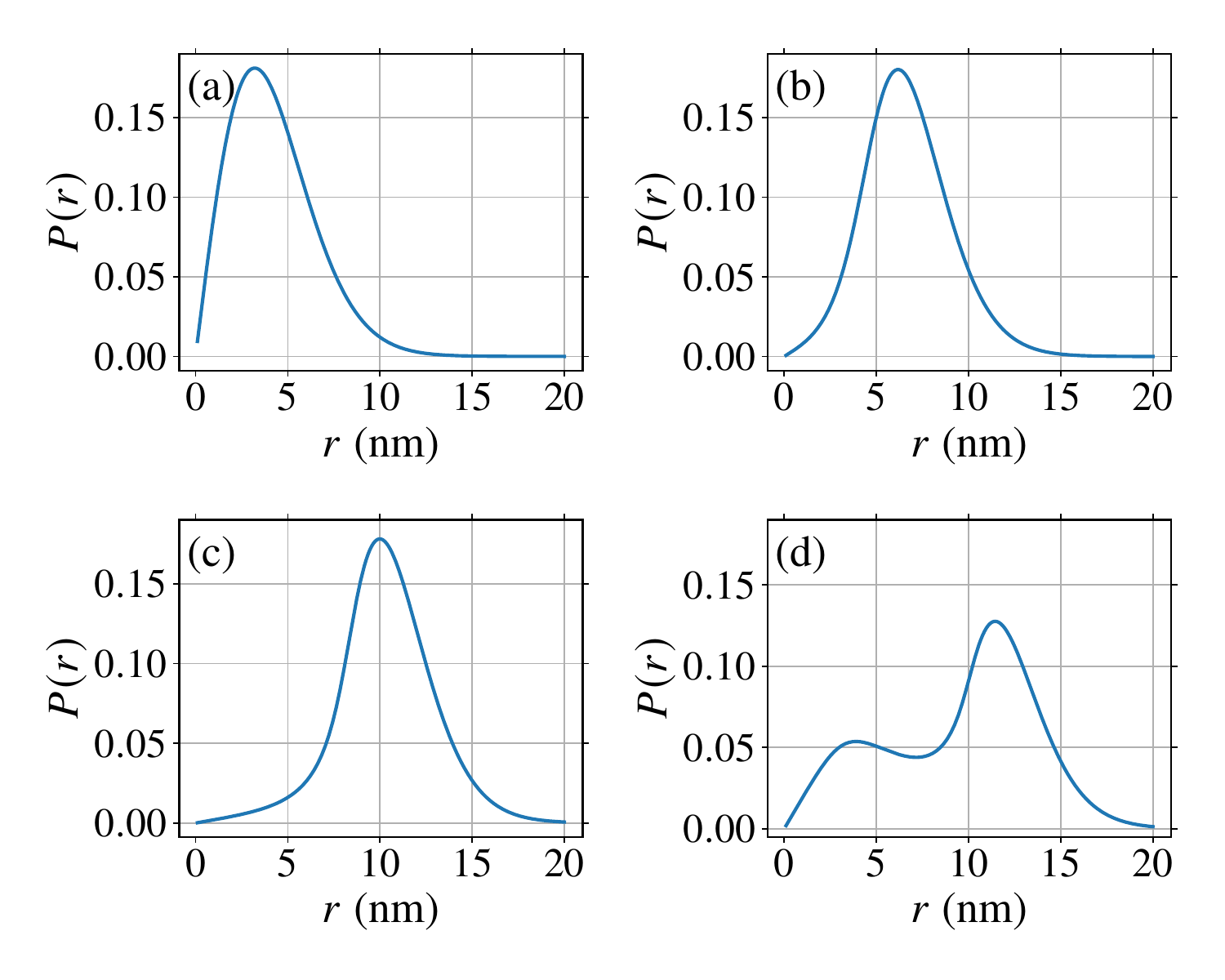}
    \caption{GS probability $P(r)=\int_0^{2\pi}d\theta\,\left|\Psi(r,\theta)\right|^2$ for the AB-stacked $\mathrm{MoTe}_2$ bilayer for (a) $p_0=1.0$ GPa, (b) $p_0=1.5$ GPa, (c) $p_0=2.5$ GPa, and (d)$p_0=3.5$ GPa. Parameters: $\alpha=10$ nm}
    \label{fig:AB_Pr_alpha10nm}
\end{figure}

Focusing on the case $\alpha=10$ nm in Fig. \ref{fig:phase_rpeak}, we observe that $r_{\rm peak}$ does not vary monotonically with increasing pressure. In particular, at $p_0=3.5$ GPa, the value of $r_{\rm peak}$ decreases sharply, indicating that the maximum of the probability density shifts back toward the center. To clarify the origin of this nonmonotonic behavior, we perform two complementary analyses. First, we examine the real-space probability-density profiles for  $p_0=1,1.5,2.5$ and $3.5$  GPa as shown in Fig. \ref{fig:AB_alpha10nm}. Interestingly, the case when $p_0=3.5$ GPa appears to have a mix of ring and dot confinement. We further plot the probability $P(r)=\int_0^{2\pi}d\theta\,\left|\Psi(r,\theta)\right|^2$ in Fig. \ref{fig:AB_Pr_alpha10nm}. For the $p_0=3.5$ GPa case, $P(r)$ has two peaks one closer to the origin and the other with maximum $P(r)$, located further away from the origin, suggesting that annular confinement is most probable in this case.

\subsection{Aharonov-Bohm oscillations in the AB-stacked ${\rm MoTe_2}$ quantum ring}
\label{sec:AB_static}

As discussed in the previous section, increasing the local pressure drives the AB-stacked bilayer from quantum-dot-like confinement into an annular regime, thereby realizing a pressure-induced quantum ring. This naturally raises the question of whether the resulting ring-like electronic states exhibit the Aharonov–Bohm effect characteristic of coherent quantum transport around a closed geometry. To investigate this, we introduce a magnetic flux $\Phi$ threaded through the center of the ring and described by the vector potential
\begin{equation}
    \vec{A}=\frac{\Phi}{2\pi r}\hat{\theta}
\end{equation}
For $r>0$, the corresponding magnetic field vanishes, while the total flux is confined to the origin,
\begin{equation}
    \vec{B}=\Phi\boldsymbol{\delta}^{(2)}(\vec{r})\hat{z}
\end{equation}
representing an idealized infinitesimally thin flux tube. The relevant magnetic flux quantum is
\begin{equation}
    \Phi_0=\frac{h}{e}
\end{equation}
where $h$ is the Planck constant, and $e>0$ is the magnitude of the elementary charge. It is natural to define the dimensionless flux
\begin{equation}
    \varphi=\Phi/\Phi_0
\end{equation}

The effect of magnetic field is incorporated into our continuum model via minimal coupling substitution
\begin{equation}
    \vec{k}\to\vec{k}+e\vec{A}
\end{equation}
which yields a Hamiltonian of the same form as in the zero-field case, but with the replacement $\nu\to \nu+\phi$. The numerical diagonalization can therefore be carried out following the same procedure described in Appendix \ref{sec:num_diag}.

\begin{figure}[h!]
    \centering
    \includegraphics[width=\linewidth]{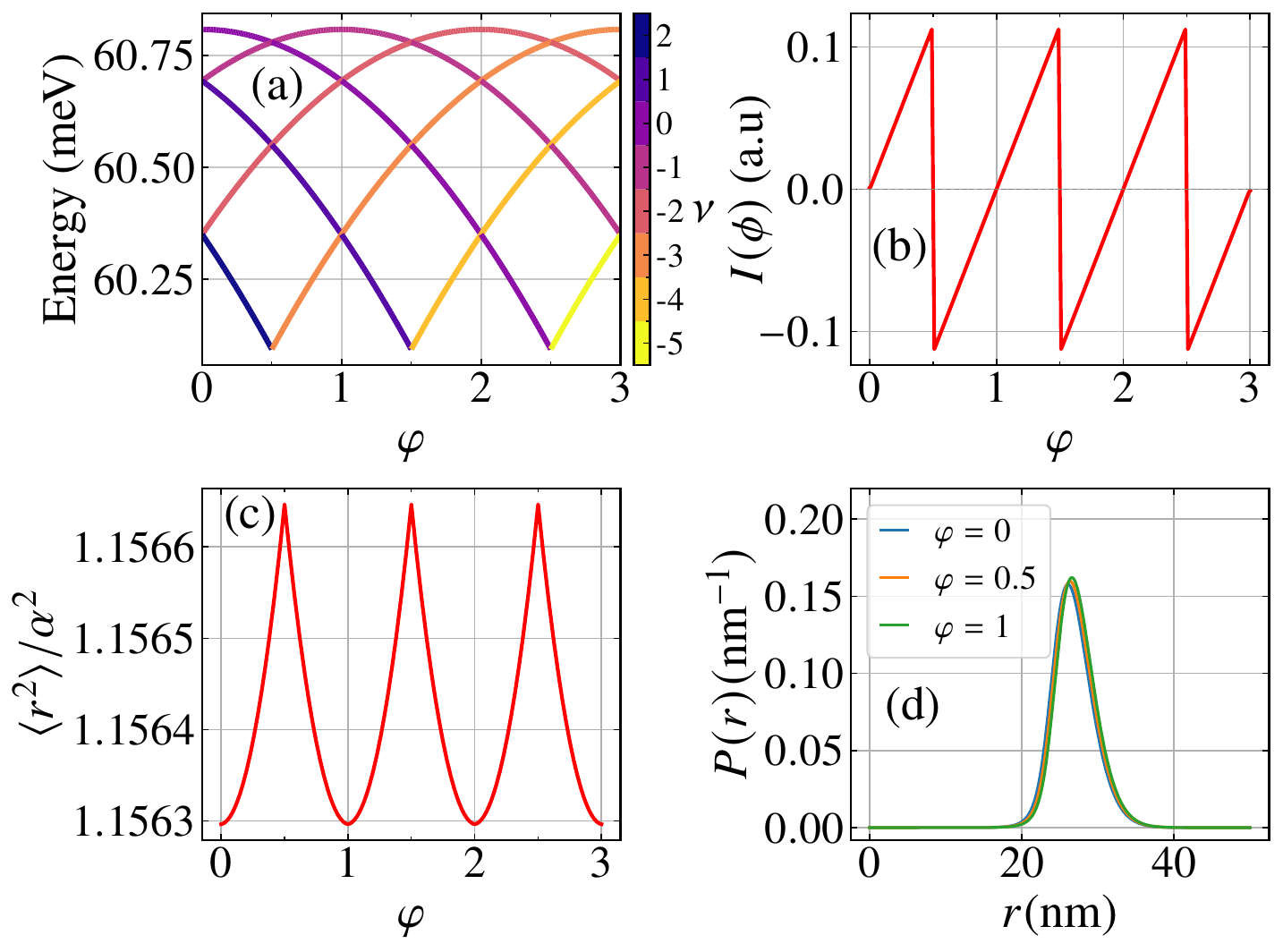}
    \caption{Flux dependence of the pressure-induced quantum ring. (a) Energies of the five lowest bound states versus reduced flux $\varphi=\Phi/\Phi_0$. Colors indicate angular momentum $\nu$; the GS exhibits half-integer AB switching ($0\to-1\to-2\to-3$), while excited states show additional integer-flux crossings due to branch reordering. (b) GS persistent current $I(\varphi)=-dE/d\varphi$, displaying the characteristic sawtooth AB oscillations. (c) Normalized radial moment $\langle r^2\rangle/\alpha^2$ as a function of flux; the periodic cusps signal a weak ``breathing'' of the finite-width ring at the angular-momentum transitions. (d) Radial probability density $P(r)$ of the GS at $\varphi=0$, $0.5$, and $1$, showing the small flux-induced displacement that underlies the breathing while preserving the overall shape. Parameters: $\alpha=25$ nm, and $p_0=3.0$ GPa.}
    \label{fig:AB_effect}
\end{figure}

As an illustration, we consider the case of $\alpha=25$ nm and $p_0=3.0$ GPa. First, we plotted the spectrum of the GS and the first four excited states as functions of the magnetic flux $\varphi$ in Fig. \ref{fig:AB_effect}(a). The GS shows multiple cusps at half integer $\varphi$ at which it changes its angular momentum quantum number $\nu$ as indicated in the color bar. To understand this, we recall that the Hamiltonian depends on $(\nu+\varphi)^2$ and transition occur when two states have the same energies, or equivalently $(\nu+\varphi)^2=(\nu'+\varphi)^2$, giving
\begin{equation}
    \varphi=-\frac{\nu+\nu'}{2}
\end{equation}
which may take integer or half-integer values depending on $\nu+\nu'$. For the ground state, the system selects the angular-momentum sector that minimizes the energy at each value of the reduced flux $\varphi$. For the excited states, the transition happens at mix of half integer and integer $\varphi$ values as shown in Fig. \ref{fig:AB_effect}(a). The resulting flux-periodic spectrum is a characteristic signature of the Aharonov–Bohm effect in a quantum ring. 

Moreover, we calculate the GS persistent current \cite{PhysRevLett.64.2074}, defined by
\begin{equation}
    I(\phi)=-\frac{1}{\Phi_0}\frac{\partial E_{\rm GS}}{\partial \varphi}
\end{equation}
and plot it as a function of the dimensionless flux $\varphi$ in Fig. \ref{fig:AB_effect}(b). The persistent current exhibits the characteristic sawtooth-like dependence on magnetic flux expected for a phase-coherent quantum ring. The discontinuous jumps occur at GS level crossings, where the angular-momentum quantum number of the GS changes as the magnetic flux is varied.

Finally, we examine how the Aharonov–Bohm flux modifies the spatial profile of the confined electronic state by plotting $\langle r^2\rangle/\alpha^2$ as a function of $\varphi$ in Fig. \ref{fig:AB_effect}(c).Interestingly, $\langle r^2\rangle/\alpha^2$ exhibits weak but systematic oscillations with the magnetic flux, a behavior absent in the ideal one-dimensional quantum-ring model, where the ring radius is fixed.  In the present pressure-induced ring, the radial degree of freedom remains finite and allows the electronic state to undergo a weak flux-induced “breathing” motion, with $\langle r^2\rangle/\alpha^2$ oscillating between $1.1563$ and $1.1567$. This behavior is also evident in the real-space probability densities shown in Fig. \ref{fig:AB_effect}(d) where the radial position of the density maximum shifts slightly as the magnetic flux is varied, consistently with the oscillations observed in Fig. \ref{fig:AB_effect}(c).

\subsection{The AA-stacked $\mathrm{MoTe}_2$ bilayer QD}
\label{sec:AA_QD}

\begin{figure}[h!]
    \centering
    \includegraphics[width=\linewidth]{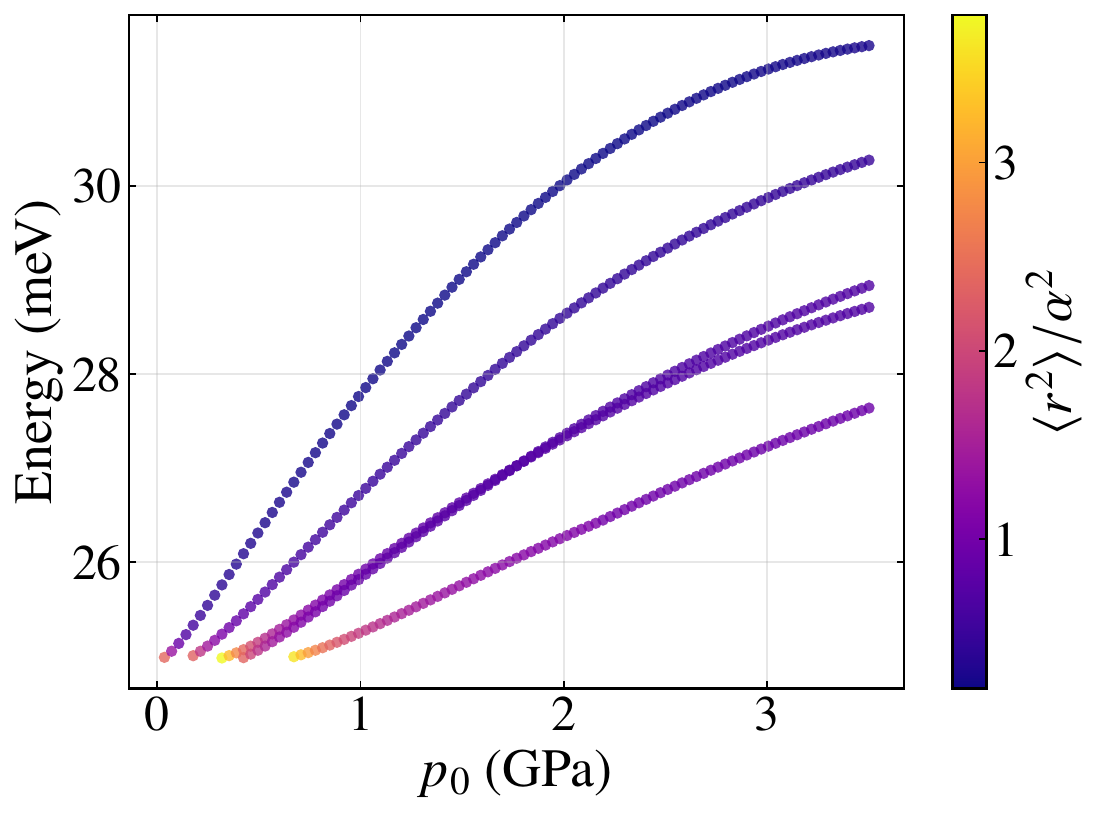}
    \caption{Five distinct localized energy branches as functions of $p_0$ for the AA-stacked $\mathrm{MoTe}_2$ bilayer. The highest-energy branch is the nondegenerate GS, while each of the other four branches is twofold degenerate. Parameters: $\alpha=25\,\mathrm{nm}$.}
    \label{fig:AA_QD}
\end{figure}

Next, we consider the AA-stacked $\mathrm{MoTe}_2$ bilayer. Figure \ref{fig:AA_QD} displays five distinct branches of localized energy levels as functions of the local pressure amplitude $p_0$ for $\alpha=25$ nm. The highest-energy branch corresponds to the nondegenerate ground state (GS), while the remaining four branches are each twofold degenerate, with only one representative of each doublet shown. The color scale represents the corresponding normalized radial extent $\langle r^2\rangle/\alpha^2$. In contrast to the AB-stacked bilayer, increasing the pressure does not induce a transition to annular confinement. Instead, the ground state (GS) remains localized near the center of the pressure profile over the entire physically accessible pressure range, retaining its predominantly quantum-dot-like character. The absence of a significant outward redistribution of the ground-state probability density highlights the crucial role of stacking configuration in governing the pressure-induced quantum dot-to-ring crossover. This behavior can also be understood from the effective potential shown in Fig. \ref{fig:effective_potentials_AB}(c). Although pressure produces a weak off-center structure in the AA-stacked case, its characteristic energy scale is much smaller than in the AB-stacked case and is insufficient to induce a substantial radial redistribution of the GS wavefunction. In particular, the pressure-induced modulation remains comparable to or smaller than the characteristic kinetic energy scale $\hbar^2/(2m^*\alpha^2)$, making the formation of an annular GS energetically unfavorable.

\subsection{Local pressure and Static confinement}
\label{sec:pressure_and_confinement}

Thus far, our calculations have neglected the static potential $U(r)$ in our calculations. In this section, we briefly discuss its influence on the confinement configurations. For the confining potential We considered a Gaussian potential $U(r)=U_0e^{-r^2/\alpha^2}$ centered at the origin. As $U_0$, increases, static confinement is expected to dominate over pressure-induced effects, driving the particle toward the center of the potential.  In the case of AB-stacked $\mathrm{MoTe}_2$ bilayer, this competition is expected to suppress the annular confinement beyond a critical value of $U_0$, restoring a predominantly quantum-dot-like state. 

\begin{figure}[h!]
    \centering
    \includegraphics[width=\linewidth]{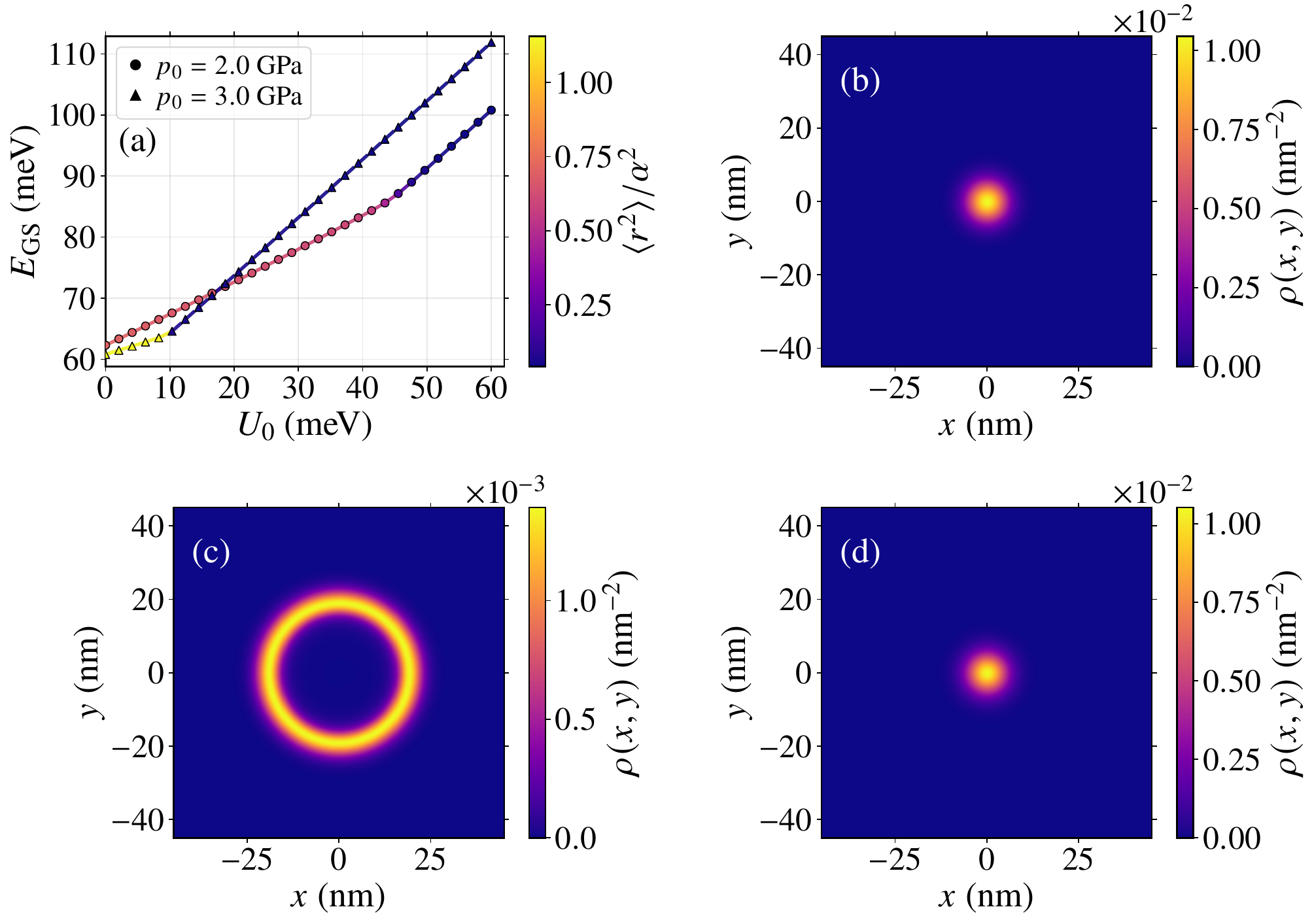}
    \caption{(a) GS energy in the AB-stacked $\mathrm{MoTe}_2$ bilayer versus $U_0$ for $p_0=2$ and $3$ GPa, where the color shows the values of $\langle r^2\rangle/\alpha^2$ reflecting the spatial extent of the wavefunction. Probability densities for $U_0=30$ meV and pressure amplitudes (b) $p_0=1.0$, (c) $p_0=2.0$, and (d) $p_0=3.0$ GPa. Parameters: $\alpha=25$ nm}
    \label{fig:AB_U0}
\end{figure}

To illustrate this behavior, we consider $\alpha=25$ nm and plot the GS energies as functions of $U_0$ at $p_0=2$ and $p_0=3$ GPa as shown in Fig. \ref{fig:AB_U0}(a). The curve colors show the values of $\langle r^2\rangle/\alpha^2$. Interestingly, the case $p_0=2$ GPa, shows a drop in the value of $\langle r^2\rangle/\alpha^2$ near $U_0=45$ meV suggesting that the annular confinement breaks down above that value. We further plot the probability densities at pressures $p_0=1,2,$ and $3$ GPa for the case $U_0=30$ meV in Figs. \ref{fig:AB_U0}(a), (b), and (c), respectively. Interestingly, the ring only appears in the $p_0=2$ GPa case while the other two cases host QD-like confinements. These findings demonstrate that the confinement geometry is governed by the competition between pressure-induced modifications of the electronic structure and the static central potential, providing an additional degree of control over the crossover between quantum-dot-like and ring-like states.

\subsection{Pressure-tunable Aharonov-Bohm Interferometer}
\label{sec:AB_interferometer}

As demonstrated in Sec. \ref{sec:dot_to_ring}, increasing the local pressure amplitude drives a crossover from quantum-dot-like to annular confinement in AB-stacked $\mathrm{MoTe}_2$ bilayers, with the onset of ring-like confinement occurring near $p_0\gtrsim1.3$ GPa. The resulting quantum-ring geometry remains well established at higher pressures, up to $p_0=3.5$ GPa, within the validity range of the pressure-dependent continuum model \cite{miftah}. The flux-dependent spectrum and persistent current discussed above demonstrate that this pressure-defined ring supports a pronounced Aharonov-Bohm response. This motivates the consideration of a two-terminal transport geometry, in which the pressure-induced quantum ring is connected to source and drain leads, enabling direct investigation of phase-coherent quantum interference through conductance measurements. As demonstrated below, the magnetic flux strongly modulates the electron transmission, establishing the proposed device as a pressure-defined, flux-tunable Aharonov–Bohm interferometer.

\begin{figure}[h!]
    \centering
    \includegraphics[width=\linewidth]{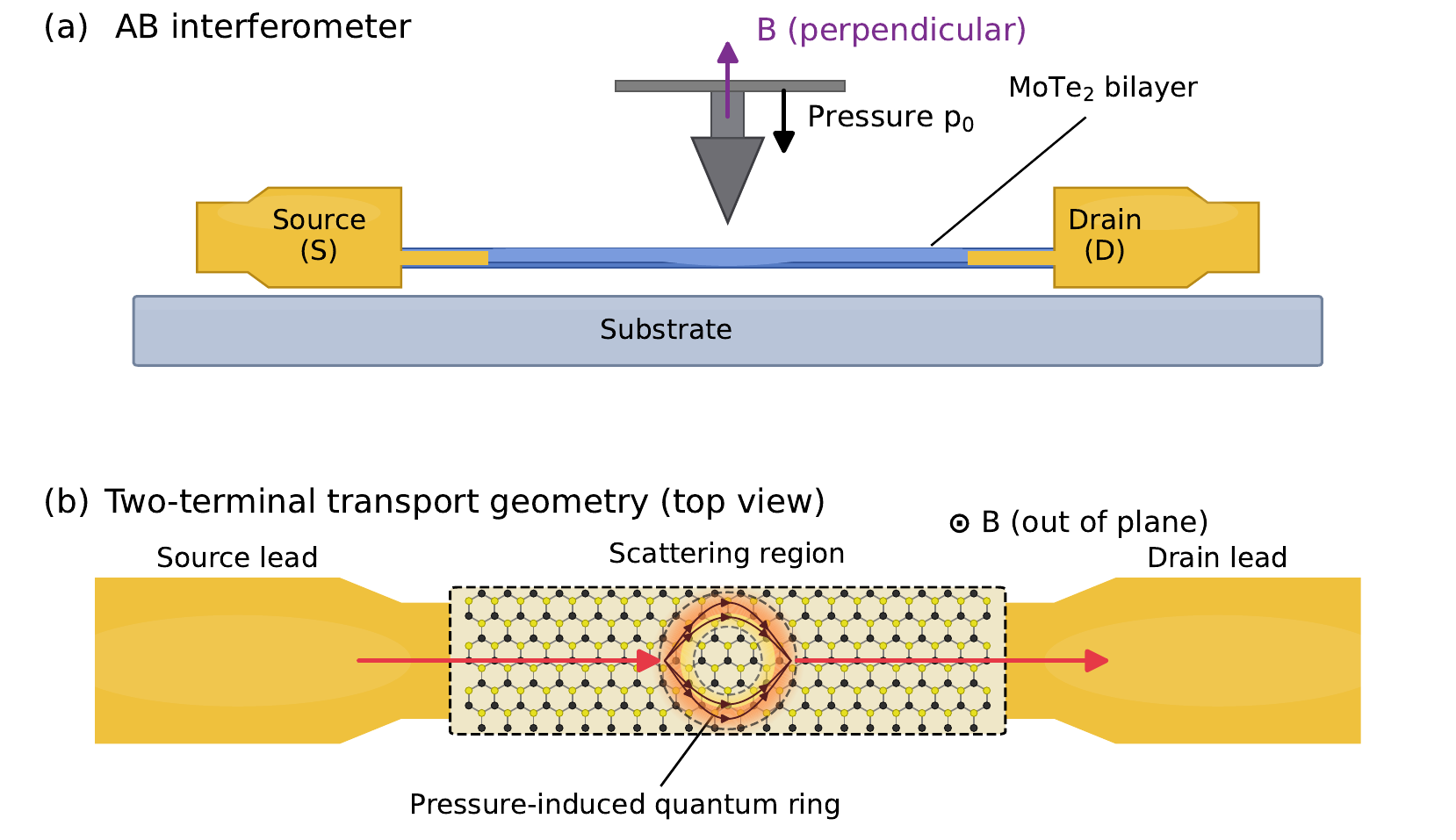}
    \caption{Pressure-tunable Aharonov-Bohm interferometer. (a) Side view of the proposed device geometry, consisting of an AB-stacked MoTe$_2$ bilayer placed on top of a substrate in close contact with source and drain leads. A local pressure probe deforms the bilayer around the center and generates the pressure-induced annular confinement, while a perpendicular magnetic field $B$ threads the device. (b) Top view of the two-terminal transport geometry. The MoTe$_2$ lattice extends continuously throughout the scattering region, while the shaded annulus represents the pressure-induced quantum-ring confinement. Electrons injected from the source split into the upper and lower arms of the ring and recombine before entering the drain, acquiring a relative Aharonov-Bohm phase determined by the enclosed magnetic flux.}
    \label{fig:device2}
\end{figure}

The proposed transport geometry is illustrated schematically in Fig. \ref{fig:device2}. The MoTe$_2$ bilayer is connected by source and drain leads, while a local pressure probe positioned above the scattering region induces annular confinement. A perpendicular magnetic field threads the pressure-defined quantum ring, introducing a relative Aharonov–Bohm phase between electrons propagating along its upper and lower arms. Electrons injected from the source can coherently traverse both paths, with their wave amplitudes recombining near the drain. The resulting constructive or destructive interference renders the electron transmission sensitive to the magnetic flux enclosed by the ring.

\begin{figure}[h!]
    \centering
    \includegraphics[width=\linewidth]{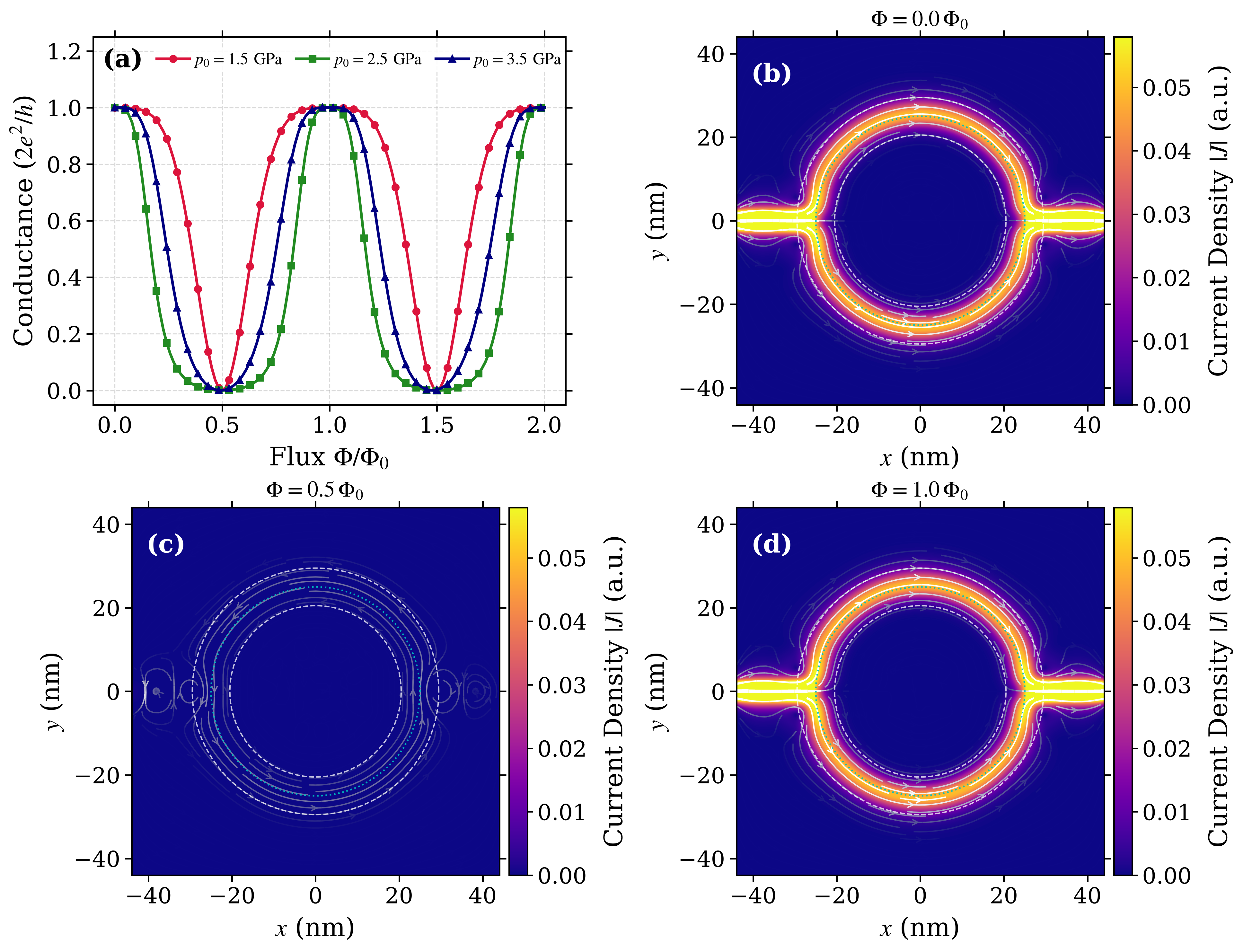}
    \caption{Phase-coherent transport and Aharonov-Bohm interferometry in the pressure-induced quantum ring device. (a) Two-terminal conductance $G$ (in units of $2e^2/h$) as a function of the dimensionless magnetic flux $\varphi=\Phi/\Phi_0$ threaded through the ring, illustrating $h/e$-periodic Aharonov-Bohm oscillations for different $p_0$. (b)-(d) Real-space current density distributions $\mathbf{J}$, for $p_0=2.5$ GPa at representative flux values of: (b) $\varphi = 0$, (c) $\varphi=0.5$ and (d) $\varphi=1.0$. The current patterns illustrate constructive transmission at integer flux and destructive interference near half-integer flux. For more details regarding current mapping, we refer to Eq.\ref{eqD1}.}
    \label{fig:transport}
\end{figure}

We next quantify the interference response of this geometry through the two-terminal conductance. Figure \ref{fig:transport}(a) shows the conductance $G$ as a function of the normalized magnetic flux $\Phi/\Phi_0$ for several values of the local pressure amplitude $p_0$. Once annular confinement is established, the conductance exhibits pronounced oscillations with a period of approximately one flux quantum, $\Phi_0=h/e$, providing a clear signature of the Aharonov–Bohm effect. Transmission is enhanced near integer values of $\Phi/\Phi_0$ and strongly suppressed near half-integer flux, reflecting the alternating constructive and destructive interference of electronic waves propagating through the two arms of the ring. The dependence of these oscillations on $p_0$ further demonstrates that the interference response can be effectively tuned through the pressure-induced confinement profile.

To visualize directly the underlying interference mechanism, we examine the corresponding real-space current distributions in Figs. \ref{fig:transport}(b)-(d). At zero flux [Fig. \ref{fig:transport}(b)], current injected from the source is distributed between the upper and lower arms of the pressure-defined ring and recombines at the drain side with a relative Aharonov-Bohm phase of zero modulo $2\pi$. The resulting constructive interference produces an enhanced transmission. At half a flux quantum [Fig. \ref{fig:transport}(c)], the two paths acquire a relative Aharonov-Bohm phase close to $\pi$, leading to destructive interference and a pronounced suppression of the transmitted current. Upon increasing the enclosed flux to one flux quantum [Fig. \ref{fig:transport}(d)], the relative phase reaches $2\pi$, restoring constructive interference and the high-transmission current pattern.

The conductance oscillations and corresponding real-space current distributions confirm that the pressure-defined annular channel supports phase-coherent Aharonov–Bohm interference within the coherent-transport regime considered here. Unlike conventional lithographically patterned quantum rings, the interferometric pathway emerges from the spatially varying electronic confinement induced by local pressure. These findings highlight the potential for realizing reconfigurable two-dimensional quantum interferometers, in which the confinement geometry is mechanically tunable while electron transmission is controlled by the enclosed magnetic flux.

\section{Conclusion}
We investigated local pressure as a means of engineering electronic confinement in the MoTe$_2$ bilayers. Using a smooth Gaussian pressure profile, we found that local pressure produces quantum-dot-like confinement in both AA-stacked and AB-stacked bilayers. However, in the AB-stacked system, increasing the pressure amplitude $p_0$ drives a crossover from conventional quantum-dot confinement to an annular regime, in which the probability density of the ground state shifts away from the center and becomes predominantly localized along a ring. In contrast, the AA-stacked bilayer remains quantum-dot-like throughout the pressure range considered, demonstrating the pronounced dependence of the pressure-induced confinement on the stacking configuration. 

The emergence of annular confinement in the AB-stacked bilayer provides a natural platform for exploring Aharonov-Bohm physics. When magnetic flux is threaded through the pressure-defined ring, the energy spectrum becomes periodic in the enclosed flux and the ground state supports a persistent current, providing characteristic signatures of the Aharonov–Bohm effect. These results demonstrate that local pressure not only modulates the confinement strength but also fundamentally alters its geometry, enabling a continuous crossover from quantum-dot-like to quantum-ring-like confinement.

We further investigate phase-coherent transport through the pressure-defined quantum ring connected to source and drain leads, using the Kwant package  \cite{Groth2014Kwant}. Electrons injected from the source propagate coherently along the upper and lower arms of the ring, and the corresponding wave amplitudes recombine near the drain. The enclosed magnetic flux modulates the relative Aharonov–Bohm phase accumulated along the two paths, leading to enhanced transmission at integer multiples of the flux quantum and pronounced suppression at half-integer multiples. The resulting flux-periodic conductance oscillations, together with the corresponding real-space current distributions, establish the pressure-defined quantum ring as a mechanically reconfigurable Aharonov–Bohm interferometer.

In summary, our findings demonstrate that local pressure provides an effective means of engineering reconfigurable quantum confinement in MoTe$_2$ bilayers. In the AB-stacked configuration, spatially varying pressure enables a continuous crossover from quantum-dot-like to quantum-ring-like confinement without the need for lithographic patterning. When connected to source and drain leads, the pressure-defined ring supports phase-coherent electron interference, with transmission controlled by the enclosed magnetic flux. These results highlight the potential of locally pressurized van der Waals bilayers for realizing mechanically reconfigurable quantum confinement and tunable Aharonov–Bohm interferometry in two-dimensional materials.

\section{Acknowledgments}
IA thanks Dr. Michael Vogl (KFUPM) for early discussions on the use of local pressure in MoTe$_2$ bilayers, which helped motivate the initial direction of this work. IA and JPFL acknowledge the support of the Natural Sciences and Engineering Research Council of Canada (NSERC) RGPIN-2022-03882 and (NRC) AQC-200-1.

\appendix
\begin{widetext}

\section{Numerical Diagonalization}
\label{sec:num_diag}
In this section, we will illustrate how to numerically diagonalize the QD Hamiltonian in Eq. \eqref{eq:H_QD}. First, we note that the Hamiltonian has a mix of momentum $\vec{k}$ and position $r$, and for proper diagonalization we need to replace $\vec{k}$ with $-i\nabla$. Secondly, if we denote the eigenstate of $H_{\rm QD}$ by $\Psi(r,\theta)=(\psi^+(r,\theta),\psi^-(r,\theta))^t$,  we find the action of the kinetic energy operator $\hat{T}$ on the state components
\begin{align}
\label{eq:pureP_App1}
    \hat{T}\psi^{\pm}(r,\theta)&= -\frac{\hbar^2}{2m_e} \nabla \cdot \left[ \frac{1}{\mu(r)} \nabla\psi^{\pm}(r,\theta) \right]\\
    &= -\frac{\hbar^2}{2m_e \mu(r)} \left[ \frac{\partial^2 \psi^\pm}{\partial r^2} + \frac{1}{r} \frac{\partial \psi^\pm}{\partial r} + \frac{1}{r^2} \frac{\partial^2 \psi^\pm}{\partial \theta^2} - \frac{1}{\mu(r)} \frac{\partial \mu}{\partial r} \frac{\partial \psi^\pm}{\partial r} \right]\nonumber 
\end{align} 
where we wrote the effective mass as $m^*(r)=m_e\mu(r)$. Next, we use the Ansatz 
\begin{equation}
\label{eqn:pureP_ansatz}
    \psi^\pm(r, \theta) =\frac{e^{i \nu \theta}}{\sqrt{2\pi}} \sqrt{\frac{\mu(r)}{r}} \chi^\pm(r)
\end{equation}
which simplifies the above expression
\begin{equation}
    \hat{T} \psi^\pm(r, \theta) = -\frac{\hbar^2}{2m_e \mu(r)} \frac{e^{i \nu \theta}}{\sqrt{2\pi}} \sqrt{\frac{\mu(r)}{r}} \left[ \frac{d^2}{dr^2} - \frac{\left( \nu^2 - \frac{1}{4} \right)}{r^2} - \frac{3}{4} \left( \frac{\mu'(r)}{\mu(r)} \right)^2 + \frac{\mu'(r)}{2r \mu(r)} + \frac{\mu''(r)}{2 \mu(r)} \right] \chi^\pm(r)
\end{equation}
where the first order derivative of the state components has been eliminated. 
\begin{equation}
    H_{\rm QD}\begin{pmatrix}
    \psi^+(r,\theta)\\
    \psi^-(r,\theta)
    \end{pmatrix}=-\frac{e^{i \nu \theta}}{\sqrt{2\pi}} \sqrt{\frac{\mu(r)}{r}}\begin{pmatrix}
 \frac{\hbar^2}{m_e \mu(r)} \hat{h}_0+ V_t(r)&-\Delta_T\\
-\Delta_T^\dagger&\frac{\hbar^2}{m_e \mu(r)} \hat{h}_0 + V_b(r)
\end{pmatrix}\begin{pmatrix}
    \chi^+(r)\\
    \chi^-(r)
    \end{pmatrix}
\end{equation}
where
\begin{equation}
\label{eqn:V_tb}
    V_\ell(r)=V_\mu(r)-U(r)-\Delta_\ell,
\end{equation}
\begin{equation}
\label{eqn:V_mu}
    V_\mu(r)=-\frac{\hbar^2}{2m_e \mu(r)} \left[  - \frac{3}{4} \left( \frac{\mu'(r)}{\mu(r)} \right)^2 + \frac{\mu'(r)}{2r \mu(r)} + \frac{\mu''(r)}{2 \mu(r)} \right],
\end{equation}
and
\begin{equation}
    \hat{h}_0=-\frac{1}{2} \frac{d^2}{dr^2} + \frac{\ell(\ell+1)}{2r^2}
\end{equation}
with $\ell=|\nu|-\frac{1}{2}$. Thus, we can rewrite the eigenvalue equation $H_{\rm QD}\Psi(r,\theta)=E\Psi(r,\theta)$ as
\begin{equation}
\label{eqn:2x2_chi_SE}
\begin{pmatrix}
\hat{h}_0+\frac{m_e \mu(r)}{\hbar^2}V_t(r)&-\frac{m_e \mu(r)}{\hbar^2}\Delta_T\\
-\frac{m_e \mu(r)}{\hbar^2}\Delta_T^\dagger&\hat{h}_0+\frac{m_e \mu(r)}{\hbar^2}V_b(r)
\end{pmatrix}\begin{pmatrix}
    \chi^+(r)\\
    \chi^-(r)
    \end{pmatrix}=-\frac{m_e \mu(r)}{\hbar^2}E\begin{pmatrix}
    \chi^+(r)\\
    \chi^-(r)
    \end{pmatrix}
\end{equation}
Now, we expand $\chi^\pm(r)$ as $\chi^\pm(r) = \sum_n C_n^\pm \varphi_n(\lambda r)$, where $C_n^\pm$ are complex expansion coefficients, $\lambda$ is a scale factor, and
\begin{equation}
    \varphi_n(y)=A_n y^{\ell+1} e^{-\frac{y^2}{2}} L_n^{\ell+\frac{1}{2}} (y^2)
\end{equation}
with $A_n = \sqrt{\frac{2\lambda\Gamma(n+1)}{\Gamma(n+\ell+\frac{3}{2})}}$ and $L_n^{\gamma}(x)$ being the (associated) Laguerre polynomial of degree $n$. One can easily verify that
\begin{equation}
    \int_0^\infty dr \, \varphi_n(\lambda r) \varphi_m(\lambda r) = \delta_{n,m}.
\end{equation}
Defining the matrix elements of the operator $\hat{\mathcal{O}}(r)$ as $\langle n |\hat{\mathcal{O}}(r)|m \rangle= \int_0^\infty dr \, \varphi_n(\lambda r) \hat{\mathcal{O}}(r) \varphi_m(\lambda r)$, we find
\begin{equation}
\langle n |\hat{h}_0|m \rangle=\frac{\lambda^2}{2} \left[ (2m + |\nu| + 1) \delta_{n,m} + \sqrt{m(m + |\nu|)} \delta_{n,m-1} + \sqrt{(m+1)(m + |\nu| + 1)} \delta_{n,m+1} \right]
\end{equation}
and for any $F(r)$, a scalar function of $r$,
\begin{equation}
 \langle n | F(r) |m \rangle=\left[\Lambda \cdot D \cdot \Lambda^T \right]_{n,m} 
\end{equation}
where $\Lambda$ is the matrix of eigenvectors of $\Omega$ whose elements are:
\begin{equation}
\Omega_{n,m} = (2m + |\nu| + 1) \delta_{n,m} - \sqrt{m(m + |\nu|)} \delta_{n,m-1} - \sqrt{(m+1)(m + |\nu| + 1)} \delta_{n,m+1}   
\end{equation}
and $D_{ks}=F(\sqrt{e_k}/\lambda)\delta_{k,s}$, with $e_k$ being the associated eigenvalue. Thus, Eq. \eqref{eqn:2x2_chi_SE} becomes
\begin{equation}
\label{eqn:2x2_chi_SE_algebraic}
\begin{pmatrix}
\boldsymbol{H}_0+\boldsymbol{W}_{t}&\boldsymbol{Q}\\
\boldsymbol{Q}^\dagger&\boldsymbol{H}_0+\boldsymbol{W}_{b}
\end{pmatrix}\begin{pmatrix}
    \boldsymbol{C}^+\\
    \boldsymbol{C}^-
    \end{pmatrix}=\tilde{E}\begin{pmatrix}
\boldsymbol{M}&\boldsymbol{0}\\
\boldsymbol{0}&\boldsymbol{M}
\end{pmatrix}\begin{pmatrix}
    \boldsymbol{C}^+\\
    \boldsymbol{C}^-
    \end{pmatrix}
\end{equation}
where
\begin{equation}
\left[\boldsymbol{H}_0\right]_{nm}=\frac{1}{\lambda^2}\langle n |\hat{h}_0|m \rangle.
\end{equation}
\begin{equation}
\left[\boldsymbol{W}_{t/b}\right]_{nm}=\frac{1}{\epsilon} \langle n | \mu(r)V_{t/b}(r) |m \rangle,
\end{equation}
\begin{equation}
\left[\boldsymbol{Q}\right]_{nm}=-\frac{1}{\epsilon} \langle n | \mu(r)\Delta_T|m \rangle,
\end{equation}
and
\begin{equation}
\left[\boldsymbol{M}\right]_{nm}=\langle n | \mu(r)|m \rangle,
\end{equation}
with $\epsilon=\hbar^2\lambda^2/m_e$ and $\tilde{E}=-E/\epsilon$. Since our units of energy is meV, we use the numerical value
\begin{equation}
    \frac{\hbar^2}{m_e}\approx 76.199642\, {\rm meV}\cdot{\rm nm^2}
\end{equation}
Finally, for a given state $\Psi(r,\theta)$, we find
\begin{equation}
    \langle r^2\rangle=\boldsymbol{C}^\dagger \mathcal{F} \boldsymbol{C}
\end{equation}
where $\boldsymbol{C}=(C^+,C^-)^t$ and
\begin{equation}
    \mathcal{F}=\begin{pmatrix}
\Lambda\cdot\mathcal{D}\cdot\Lambda^T&\boldsymbol{0}\\
\boldsymbol{0}&\Lambda\cdot\mathcal{D}\cdot\Lambda^T
\end{pmatrix}
\end{equation}
where $\mathcal{D}_{ks}=e_k \mu(\sqrt{e_k}/\lambda)/\lambda^2\delta_{ks}$. The eigenvalue problem is of the form
\begin{equation}
    A\boldsymbol{C}=\tilde{E}B\boldsymbol{C}
\end{equation}
where
\begin{equation}
    A=\begin{pmatrix}
\boldsymbol{H}_0+\boldsymbol{W}_{t}&\boldsymbol{Q}\\
\boldsymbol{Q}^\dagger&\boldsymbol{H}_0+\boldsymbol{W}_{b}
\end{pmatrix},\quad B=\begin{pmatrix}
\boldsymbol{M}&\boldsymbol{0}\\
\boldsymbol{0}&\boldsymbol{M}
\end{pmatrix}
\end{equation}


\section{Real-space tight-binding model and quantum transport}
To investigate quantum transport across the pressure-induced ring in AB-stacked $\mathrm{MoTe}_2$ bilayers, we define the low-energy effective model on a real space. The Hamiltonian of the discretized adopted model is given by

\begin{equation}  {\label{eq_B1}} 
H = \sum_i \left(\varepsilon_i(p_0)-\frac{4\hbar^2}{2a_0m^*(p_0)}  \right)  c_i^\dagger c_i - \sum_{\langle i,j\rangle} t_{ij}(p_0)c_i^\dagger c_j+\mathrm{h.c.} . 
\end{equation}

Where, $c_i^\dagger$ and $c_i$ are the creation and annihilation operators at site $\vec{r}_i=(x_i,y_i)$, respectively, and $\langle i,j\rangle$ denotes nearest-neighbor pairs. The hopping parameter $t_{ij}$ is determined by the pressure-dependent effective mass $m^*(p_0)$ introduced in Eq.\ref{eq_2} and $a_0=0.3487 nm$. The corresponding onsite energy $\varepsilon_i$ takes the form
\begin{equation} 
\varepsilon(\vec{r}, p_0) = V_{\mathrm{ring}}(\vec{r}, p_0) +V_{\mathrm{tip}}(\vec{r}, p_0)-U_0, 
\end{equation}

where $V_{\mathrm{ring}}$ describes the ring confinement, $V_{\mathrm{tip}}$ provides the tip repulsion potential, and
$U_0$ represents the global electrostatic gate offset. Here, $\vec{r}=\sqrt{x^2+y^2}$ is the radial distance from the center of the pressure tip.

As discussed in the main text, the pressure-induced spatial modulation of the intralayer hopping $\Delta_l(\vec{r})$ gives rise to an annular confinement minimum at $R_{\mathrm{ring}}$. We describe this channel using an attractive Gaussian potential,

\begin{equation} {\label{eq_B3}}
V_{\mathrm{ring}}(\vec{r}, p_0)=-V_{\mathrm{0}}(p_0)\exp\left[-\left(\frac{\vec{r}-R_{\mathrm{ring}}}{w_{\mathrm{ring}}}\right)^2\right],
\end{equation}
where $V_{\mathrm{0}}$ is the confinement strength and it is defined to be proportional to the applied pressure $p_0$, and $w_{\mathrm{ring}}$
characterizes the radial half-width of the ring channel. The tip induced potential effects is taken  as a weak harmonic potential,
\begin{equation}
V_{\mathrm{tip}}(\vec{r})=\frac{1}{2}\kappa \vec{r}^2,
\qquad
\kappa=\frac{2\delta\varepsilon}{R_{\mathrm{ring}}^2},
\end{equation}
where $\delta\varepsilon\ll V_{\mathrm{0}}$ ensures that the potential varies only weakly across the active ring and provides a smooth confinement without introducing artificial hard-wall effects.

The orbital coupling to an out-of-plane magnetic flux $\Phi$ is introduced through the Aharonov--Bohm vector potential $\mathbf{A}(\vec{r})=
\frac{\Phi}{2\pi r}\hat{\boldsymbol{\theta}}$ \cite{AharonovBohm1959}, as defined in Eq.~(12) of Section~III.B. This describes an idealized flux line localized at the origin, with $\mathbf{B}=\Phi\,\delta(\vec{r})\hat{\mathbf{z}}$. 
For numerical calculations on the discrete lattice, we regularize the vector potential in Cartesian coordinates as
\begin{equation}
A_x(x,y)=
-\frac{\Phi}{2\pi}\frac{y}{r^2+\eta^2},
\qquad
A_y(x,y)=
\frac{\Phi}{2\pi}\frac{x}{r^2+\eta^2},
\end{equation}
where $\eta\approx1.0\,\mathrm{nm}$ regularizes the vector potential near the origin. The magnetic field enters the lattice Hamiltonian through the Peierls substitution,
\begin{equation}
t_{ij}=t(p_0)\exp(-i\theta_{ij}),
\qquad
\theta_{ij}=
\frac{e}{\hbar}
\int_{\vec{r}_j}^{\vec{r}_i}
\mathbf{A}(\vec{r})\cdot d\vec{r},
\end{equation}
where $e$ is the magnitude of the electron charge. Using the
midpoint approximation to evaluate the line integral along each
nearest-neighbor bond gives
\begin{equation}
\theta_{i,i+\hat{\mathbf{x}}}   
=
-2\pi\left(\frac{\Phi}{\Phi_0}\right)
\frac{\bar{y}a}{2\pi(\bar{r}^2+\eta^2)},
\qquad
\theta_{i,i+\hat{\mathbf{y}}}
=
2\pi\left(\frac{\Phi}{\Phi_0}\right)
\frac{\bar{x}a}{2\pi(\bar{r}^2+\eta^2)},
\end{equation}
where
\begin{equation}
\bar{x}=\frac{x_i+x_j}{2},
\qquad
\bar{y}=\frac{y_i+y_j}{2},
\qquad
\bar{r}^2=\bar{x}^2+\bar{y}^2,
\end{equation}
and $\Phi_0=h/e=2\pi\hbar/e$ is the magnetic flux quantum. The corresponding directional hopping amplitudes are
\begin{equation}
t_{i,i+\hat{\mathbf{x}}}
=
-t(p_0)\exp\left[
i\,2\pi\left(\frac{\Phi}{\Phi_0}\right)
\frac{\bar{y}a}{2\pi(\bar{r}^2+\eta^2)}
\right],
\end{equation}
and
\begin{equation}
t_{i,i+\hat{\mathbf{y}}}
=
-t(p_0)\exp\left[
-i\,2\pi\left(\frac{\Phi}{\Phi_0}\right)
\frac{\bar{x}a}{2\pi(\bar{r}^2+\eta^2)}
\right].
\end{equation}
The hopping amplitudes satisfy the Hermiticity condition $t_{ji}=t_{ij}^*$, since reversing the bond direction reverses the sign of the Peierls phase.

The scattering region is connected along the $x$ direction to two semi-infinite metallic leads, which serve as electron reservoirs. The hopping parameter in the leads is independent of the pressure applied to the central region. The lead parameters are chosen to ensure that propagating modes are available at the zero, Fermi energy, zero temperature and in the linear-response regime, the two-terminal conductance is calculated using the Landauer--Büttiker formula \cite{Landauer1957,Buttiker1986}:
\begin{equation}
G=
\frac{2e^2}{h}\mathcal{T}(E_F) = \frac{2e^2}{h} \operatorname{Tr}\left[\Gamma_D(E_F)G^R(E_F)\Gamma_S(E_F)G^A(E_F)\right].
\end{equation}
The corresponding broadening matrices are defined as
\begin{equation}
\Gamma_{\alpha}(E_F)= i\left[\Sigma_{\alpha}^R(E_F)-\Sigma_{\alpha}^A(E_F)\right],  
\end{equation}
where $\alpha\in\{S,D\}$ denotes the source or drain lead. Alternatively, the total transmission can be expressed in terms of the transmission block $\mathbf{t}(E_F)$ of the scattering matrix as
\begin{equation}
\mathcal{T}(E_F)=\operatorname{Tr}\left[t^\dagger(E_F)t(E_F)\right] = \sum_{n,m}|t_{nm}(E_F)|^2.
\end{equation}
The transmission and conductance are calculated numerically using the recursive Green's-function method implemented in Kwant package \cite{Groth2014Kwant}.

To map the  current  in real space, we calculate the local current by collecting the scattering states \cite{Groth2014Kwant}, using the 
\begin{equation}\label{eqD1}
J_{}^{ij}=\Phi^{*}\left(i \sum_{\gamma}^{}H^{*}_{ij\gamma} H_{i\gamma}-H_{i\gamma}H_{ij\gamma} \right) \Phi,
\end{equation}

\noindent where $H_{ij}$ is a matrix with zero elements except for those connecting the sites $i$ and $j$. In this case, the hopping matrices are pressure dependent and obtained from the second term of Eq. \ref{eq_B1}. 
The current mapping $J^{i, j}$ in Eq. \ref{eqD1} is derived from the continuity equation using the density operator $\rho_q$  and expressed as \cite{Nikolic2006}
\begin{equation}\label{eqD2}
\rho_q=\sum_{i}\Phi^{*}_{i} H_{q}=0, \qquad \frac{\partial \rho_q}{\partial t}-\sum_{j}^{}J^{i, j}=0.
\end{equation}

where $\Phi$ is the wave eigenstates of the propagating modes of the source lead, as shown in Fig. \ref{fig:device2}. Here $q$ defines all sites or hoppings in the scattering region.  $H_{ij}$ is a matrix with zero elements except for those connecting the sites a and b. 

\newpage

\end{widetext}

\bibliography{ref.bib}

@article{Schaibley2016-xq,
  title     = {Valleytronics in {2D} materials},
  author    = {Schaibley, John R. and Yu, Hongyi and Clark, Genevieve and Rivera, Pasqual and Ross, Jason S. and Seyler, Kyle L. and Yao, Wang and Xu, Xiaodong},
  journal   = {Nat. Rev. Mater.},
  publisher = {Springer Science and Business Media LLC},
  volume    = {1},
  number    = {11},
  pages     = {16055},
  month     = aug,
  year      = {2016},
  doi       = {10.1038/natrevmats.2016.55},
  url       = {https://doi.org/10.1038/natrevmats.2016.55}
}

@article{Kim2024-gz,
  title     = {Confinement of excited states in two-dimensional, in-plane, quantum heterostructures},
  author    = {Kim, Gwangwoo and Huet, Benjamin and Stevens, Christopher E. and Jo, Kiyoung and Tsai, Jeng-Yuan and Bachu, Saiphaneendra and Leger, Meghan and Song, Seunguk and Rahaman, Mahfujur and Ma, Kyung Yeol and Glavin, Nicholas R. and Shin, Hyeon Suk and Alem, Nasim and Yan, Qimin and Hendrickson, Joshua R. and Redwing, Joan M. and Jariwala, Deep},
  journal   = {Nat. Commun.},
  publisher = {Springer Science and Business Media LLC},
  volume    = {15},
  number    = {1},
  pages     = {6361},
  month     = jul,
  year      = {2024},
  doi       = {10.1038/s41467-024-50653-x},
  url       = {https://doi.org/10.1038/s41467-024-50653-x}
}

@article{Kim2024-nw,
  title     = {The future of two-dimensional semiconductors beyond {Moore's} law},
  author    = {Kim, Ki Seok and Kwon, Junyoung and Ryu, Huije and Kim, Changhyun and Kim, Hyunseok and Lee, Eun-Kyu and Lee, Doyoon and Seo, Seunghwan and Han, Ne Myo and Suh, Jun Min and Kim, Jekyung and Song, Min-Kyu and Lee, Sangho and Seol, Minsu and Kim, Jeehwan},
  journal   = {Nat. Nanotechnol.},
  publisher = {Springer Science and Business Media LLC},
  volume    = {19},
  number    = {7},
  pages     = {895--906},
  month     = jul,
  year      = {2024},
  doi       = {10.1038/s41565-024-01695-1},
  url       = {https://doi.org/10.1038/s41565-024-01695-1}
}

@article{Kouwenhoven2001-di,
  title     = {Few-electron quantum dots},
  author    = {Kouwenhoven, L. P. and Austing, D. G. and Tarucha, S.},
  journal   = {Rep. Prog. Phys.},
  publisher = {IOP Publishing},
  volume    = {64},
  number    = {6},
  pages     = {701--736},
  month     = jun,
  year      = {2001},
  doi       = {10.1088/0034-4885/64/6/201},
  url       = {https://doi.org/10.1088/0034-4885/64/6/201}
}

@article{RevModPhys.79.1217,
  title     = {Spins in few-electron quantum dots},
  author    = {Hanson, R. and Kouwenhoven, L. P. and Petta, J. R. and Tarucha, S. and Vandersypen, L. M. K.},
  journal   = {Rev. Mod. Phys.},
  volume    = {79},
  issue     = {4},
  pages     = {1217--1265},
  numpages  = {49},
  year      = {2007},
  month     = oct,
  publisher = {American Physical Society},
  doi       = {10.1103/RevModPhys.79.1217},
  url       = {https://doi.org/10.1103/RevModPhys.79.1217}
}

@article{PhysRevLett.105.067404,
  title     = {Effect of Mechanical Strain on the Optical Properties of Quantum Dots: Controlling Exciton Shape, Orientation, and Phase with a Mechanical Strain},
  author    = {Bryant, Garnett W. and Zieli{\'{n}}ski, M. and Malkova, Natalia and Sims, James and Jask{\'o}lski, W. and Aizpurua, Javier},
  journal   = {Phys. Rev. Lett.},
  volume    = {105},
  issue     = {6},
  pages     = {067404},
  numpages  = {4},
  year      = {2010},
  month     = aug,
  publisher = {American Physical Society},
  doi       = {10.1103/PhysRevLett.105.067404},
  url       = {https://doi.org/10.1103/PhysRevLett.105.067404}
}

@article{Geim2013-nz,
  title     = {Van der Waals heterostructures},
  author    = {Geim, A. K. and Grigorieva, I. V.},
  journal   = {Nature},
  publisher = {Springer Science and Business Media LLC},
  volume    = {499},
  number    = {7459},
  pages     = {419--425},
  month     = jul,
  year      = {2013},
  doi       = {10.1038/nature12385},
  url       = {https://doi.org/10.1038/nature12385}
}

@article{Novoselov2016-yl,
  title     = {{2D} materials and van der Waals heterostructures},
  author    = {Novoselov, K. S. and Mishchenko, A. and Carvalho, A. and {Castro Neto}, A. H.},
  journal   = {Science},
  publisher = {American Association for the Advancement of Science (AAAS)},
  volume    = {353},
  number    = {6298},
  pages     = {aac9439},
  month     = jul,
  year      = {2016},
  doi       = {10.1126/science.aac9439},
  url       = {https://doi.org/10.1126/science.aac9439}
}

@article{Liu2016-ea,
  title     = {Van der Waals heterostructures and devices},
  author    = {Liu, Yuan and Weiss, Nathan O. and Duan, Xidong and Cheng, Hung-Chieh and Huang, Yu and Duan, Xiangfeng},
  journal   = {Nat. Rev. Mater.},
  publisher = {Springer Science and Business Media LLC},
  volume    = {1},
  number    = {9},
  pages     = {16042},
  month     = jul,
  year      = {2016},
  doi       = {10.1038/natrevmats.2016.42},
  url       = {https://doi.org/10.1038/natrevmats.2016.42}
}

@article{PhysRevB.97.195454,
  title = {Theory of strain-induced confinement in transition metal dichalcogenide monolayers},
  author = {Brooks, Matthew and Burkard, Guido},
  journal = {Phys. Rev. B},
  volume = {97},
  issue = {19},
  pages = {195454},
  numpages = {8},
  year = {2018},
  month = {May},
  publisher = {American Physical Society},
  doi = {10.1103/PhysRevB.97.195454},
  url = {https://link.aps.org/doi/10.1103/PhysRevB.97.195454}
}

@article{Chirolli2019-ze,
  title     = {Strain-induced bound states in transition-metal dichalcogenide bubbles},
  author    = {Chirolli, L. and Prada, E. and Guinea, F. and Rold{\'a}n, R. and San-Jose, P.},
  journal   = {2D Mater.},
  publisher = {IOP Publishing},
  volume    = {6},
  number    = {2},
  pages     = {025010},
  month     = apr,
  year      = {2019},
  doi       = {10.1088/2053-1583/ab0113},
  url       = {https://doi.org/10.1088/2053-1583/ab0113}
}

@article{Xu2014-aa,
  title     = {Spin and pseudospins in layered transition metal dichalcogenides},
  author    = {Xu, Xiaodong and Yao, Wang and Xiao, Di and Heinz, Tony F.},
  journal   = {Nat. Phys.},
  publisher = {Springer Science and Business Media LLC},
  volume    = {10},
  number    = {5},
  pages     = {343--350},
  month     = may,
  year      = {2014},
  doi       = {10.1038/nphys2942},
  url       = {https://doi.org/10.1038/nphys2942}
}

@article{PhysRevB.110.115410,
  title = {Electronic band structure of high-symmetry homobilayers of transition metal dichalcogenides},
  author = {Zhang, Heng and Xu, Wen and Xiao, Yiming and Peeters, Francois M. and Milo{\ifmmode \check{s}\else \v{s}\fi{}}evi{\ifmmode \acute{c}\else \'{c}\fi{}}, Milorad V.},
  journal = {Phys. Rev. B},
  volume = {110},
  issue = {11},
  pages = {115410},
  numpages = {13},
  year = {2024},
  month = {Sep},
  publisher = {American Physical Society},
  doi = {10.1103/PhysRevB.110.115410},
  url = {https://link.aps.org/doi/10.1103/PhysRevB.110.115410}
}

@article{Liu2015-md,
  title     = {Electronic structures and theoretical modelling of two-dimensional {group-VIB} transition metal dichalcogenides},
  author    = {Liu, Gui-Bin and Xiao, Di and Yao, Yugui and Xu, Xiaodong and Yao, Wang},
  journal   = {Chem. Soc. Rev.},
  publisher = {Royal Society of Chemistry (RSC)},
  volume    = {44},
  number    = {9},
  pages     = {2643--2663},
  month     = may,
  year      = {2015},
  doi       = {10.1039/C4CS00301B},
  url       = {https://doi.org/10.1039/C4CS00301B}
}

@article{PhysRevB.89.075409,
  title = {Stacking effects on the electronic and optical properties of bilayer transition metal dichalcogenides ${\mathrm{MoS}}_{2}$, ${\mathrm{MoSe}}_{2}$, ${\mathrm{WS}}_{2}$, and ${\mathrm{WSe}}_{2}$},
  author = {He, Jiangang and Hummer, Kerstin and Franchini, Cesare},
  journal = {Phys. Rev. B},
  volume = {89},
  issue = {7},
  pages = {075409},
  numpages = {11},
  year = {2014},
  month = {Feb},
  publisher = {American Physical Society},
  doi = {10.1103/PhysRevB.89.075409},
  url = {https://link.aps.org/doi/10.1103/PhysRevB.89.075409}
}

@article{PhysRevB.102.235418,
  title = {Tuning band gaps in twisted bilayer $\mathrm{Mo}{\mathrm{S}}_{2}$},
  author = {Zhang, Yipei and Zhan, Zhen and Guinea, Francisco and Silva-Guill{\'e}n, Jose {\'A}ngel and Yuan, Shengjun},
  journal = {Phys. Rev. B},
  volume = {102},
  issue = {23},
  pages = {235418},
  numpages = {7},
  year = {2020},
  month = {Dec},
  publisher = {American Physical Society},
  doi = {10.1103/PhysRevB.102.235418},
  url = {https://link.aps.org/doi/10.1103/PhysRevB.102.235418}
}

@article{PhysRevLett.122.086402,
  title = {Topological Insulators in Twisted Transition Metal Dichalcogenide Homobilayers},
  author = {Wu, Fengcheng and Lovorn, Timothy and Tutuc, Emanuel and Martin, Ivar and MacDonald, A. H.},
  journal = {Phys. Rev. Lett.},
  volume = {122},
  issue = {8},
  pages = {086402},
  numpages = {5},
  year = {2019},
  month = {Feb},
  publisher = {American Physical Society},
  doi = {10.1103/PhysRevLett.122.086402},
  url = {https://link.aps.org/doi/10.1103/PhysRevLett.122.086402}
}

@article{PhysRevB.105.L081108,
  title = {Moir\'e induced topology and flat bands in twisted bilayer ${\mathrm{WSe}}_{2}$: A first-principles study},
  author = {Kundu, Sudipta and Naik, Mit H. and Krishnamurthy, H. R. and Jain, Manish},
  journal = {Phys. Rev. B},
  volume = {105},
  issue = {8},
  pages = {L081108},
  numpages = {6},
  year = {2022},
  month = {Feb},
  publisher = {American Physical Society},
  doi = {10.1103/PhysRevB.105.L081108},
  url = {https://link.aps.org/doi/10.1103/PhysRevB.105.L081108}
}

@article{PhysRevLett.132.096602,
  title = {Magic Angles and Fractional Chern Insulators in Twisted Homobilayer Transition Metal Dichalcogenides},
  author = {Morales-Dur{\'a}n, Nicol{\'a}s and Wei, Nemin and Shi, Jingtian and MacDonald, Allan H.},
  journal = {Phys. Rev. Lett.},
  volume = {132},
  issue = {9},
  pages = {096602},
  numpages = {6},
  year = {2024},
  month = {Mar},
  publisher = {American Physical Society},
  doi = {10.1103/PhysRevLett.132.096602},
  url = {https://link.aps.org/doi/10.1103/PhysRevLett.132.096602}
}

@article{PhysRevB.99.024111,
  title = {Pressure effect on the electronic, structural, and vibrational properties of layered $2H\ensuremath{-}{\mathrm{MoTe}}_{2}$},
  author = {Zhao, Xiao-Miao and Liu, Han-yu and Goncharov, Alexander F. and Zhao, Zhi-Wei and Struzhkin, Viktor V. and Mao, Ho-Kwang and Gavriliuk, Alexander G. and Chen, Xiao-Jia},
  journal = {Phys. Rev. B},
  volume = {99},
  issue = {2},
  pages = {024111},
  numpages = {6},
  year = {2019},
  month = {Jan},
  publisher = {American Physical Society},
  doi = {10.1103/PhysRevB.99.024111},
  url = {https://link.aps.org/doi/10.1103/PhysRevB.99.024111}
}

@article{th65-gff6,
  title = {Hydrostatic Pressure-Enhanced Correlated Magnetism and Chern Insulator in Moir\'e ${\mathrm{WSe}}_{2}$},
  author = {Jiao, Pengfei and Qian, Chenghao and Mao, Ning and Chang, Xumin and Xiao, Jiayong and Liu, Feng and Wang, Shaozheng and Wu, Xiaokai and Peng, Di and Xu, Cheng and Dong, Hongliang and Zheng, Yuchen and Wu, Juncai and Zheng, Tong and Watanabe, Kenji and Taniguchi, Takashi and Jia, Jinfeng and Liu, Xiaoxue and Shi, Zhiwen and Wang, Shiyong and Chen, Guorui and Li, Tingxin and Zhong, Ruidan and Zhang, Yang and Qian, Dong and Chen, Zhiqiang and Jiang, Shengwei},
  journal = {Phys. Rev. X},
  volume = {16},
  issue = {1},
  pages = {011068},
  numpages = {10},
  year = {2026},
  month = {Mar},
  publisher = {American Physical Society},
  doi = {10.1103/th65-gff6},
  url = {https://link.aps.org/doi/10.1103/th65-gff6}
}

@article{4742-rldh,
  title = {Pressure-tunable generalized Wigner crystal and fractional Chern insulator in twisted ${\mathrm{MoTe}}_{2}$},
  author = {Wang, Bingbing and Yu, Junxi and Liu, Cheng-Cheng},
  journal = {Phys. Rev. B},
  volume = {113},
  issue = {20},
  pages = {L201409},
  numpages = {7},
  year = {2026},
  month = {May},
  publisher = {American Physical Society},
  doi = {10.1103/4742-rldh},
  url = {https://link.aps.org/doi/10.1103/4742-rldh}
}

@article{Kim2026-sq,
  title     = {Precision local strain engineering in {2D} semiconductors and their van der {Waals} heterostructures},
  author    = {Kim, Byeong Chan and Kim, Yoona and Lee, Gwan-Hyoung},
  journal   = {Small},
  publisher = {Wiley},
  volume    = {22},
  number    = {20},
  pages     = {e12902},
  month     = apr,
  year      = {2026},
  doi       = {10.1002/smll.202512902},
  url       = {https://doi.org/10.1002/smll.202512902}
}

@article{Branny2017-nz,
  title     = {Deterministic strain-induced arrays of quantum emitters in a two-dimensional semiconductor},
  author    = {Branny, Artur and Kumar, Santosh and Proux, Rapha{\"e}l and Gerardot, Brian D.},
  journal   = {Nat. Commun.},
  publisher = {Springer Science and Business Media LLC},
  volume    = {8},
  number    = {1},
  pages     = {15053},
  month     = may,
  year      = {2017},
  doi       = {10.1038/ncomms15053},
  url       = {https://doi.org/10.1038/ncomms15053}
}

@article{RevModPhys.95.025003,
  title = {Semiconductor spin qubits},
  author = {Burkard, Guido and Ladd, Thaddeus D. and Pan, Andrew and Nichol, John M. and Petta, Jason R.},
  journal = {Rev. Mod. Phys.},
  volume = {95},
  issue = {2},
  pages = {025003},
  numpages = {58},
  year = {2023},
  month = {Jun},
  publisher = {American Physical Society},
  doi = {10.1103/RevModPhys.95.025003},
  url = {https://link.aps.org/doi/10.1103/RevModPhys.95.025003}
}

@article{PhysRevApplied.13.054058,
  title = {Gate-Defined Accumulation-Mode Quantum Dots in Monolayer and Bilayer ${\mathrm{W}\mathrm{Se}}_{2}$},
  author = {Davari, S. and Stacy, J. and Mercado, A.M. and Tull, J.D. and Basnet, R. and Pandey, K. and Watanabe, K. and Taniguchi, T. and Hu, J. and Churchill, H.O.H.},
  journal = {Phys. Rev. Appl.},
  volume = {13},
  issue = {5},
  pages = {054058},
  numpages = {8},
  year = {2020},
  month = {May},
  publisher = {American Physical Society},
  doi = {10.1103/PhysRevApplied.13.054058},
  url = {https://link.aps.org/doi/10.1103/PhysRevApplied.13.054058}
}

@article{M_C_de_Melo2021-kq,
  title     = {Optical signatures of defect centers in transition metal dichalcogenide monolayers},
  author    = {M. C. de Melo, Pedro Miguel and Zanolli, Zeila and Verstraete, Matthieu J.},
  journal   = {Adv. Quantum Technol.},
  publisher = {Wiley},
  volume    = {4},
  number    = {3},
  pages     = {2000118},
  month     = mar,
  year      = {2021},
  doi       = {10.1002/qute.202000118},
  url       = {https://doi.org/10.1002/qute.202000118}
}

@article{Dirnberger2021-uc,
  title     = {{Quasi-1D} exciton channels in strain-engineered {2D} materials},
  author    = {Dirnberger, Florian and Ziegler, Jonas D. and {Faria Junior}, Paulo E. and Bushati, Rezlind and Taniguchi, Takashi and Watanabe, Kenji and Fabian, Jaroslav and Bougeard, Dominique and Chernikov, Alexey and Menon, Vinod M.},
  journal   = {Sci. Adv.},
  publisher = {American Association for the Advancement of Science (AAAS)},
  volume    = {7},
  number    = {44},
  pages     = {eabj3066},
  month     = oct,
  year      = {2021},
  doi       = {10.1126/sciadv.abj3066},
  url       = {https://doi.org/10.1126/sciadv.abj3066}
}

@article{Koo2024-qu,
  title     = {Dynamical control of nanoscale light-matter interactions in low-dimensional quantum materials},
  author    = {Koo, Yeonjeong and Moon, Taeyoung and Kang, Mingu and Joo, Huitae and Lee, Changjoo and Lee, Hyeongwoo and Kravtsov, Vasily and Park, Kyoung-Duck},
  journal   = {Light Sci. Appl.},
  publisher = {Springer Science and Business Media LLC},
  volume    = {13},
  number    = {1},
  pages     = {30},
  month     = jan,
  year      = {2024},
  doi       = {10.1038/s41377-024-01380-x},
  url       = {https://doi.org/10.1038/s41377-024-01380-x}
}

@article{Ai2025-fe,
  title     = {Local strain engineering of two-dimensional transition metal dichalcogenides towards quantum emitters},
  author    = {Ai, Ruoqi and Cui, Ximin and Li, Yang and Zhuo, Xiaolu},
  journal   = {Nano-Micro Lett.},
  publisher = {Springer Science and Business Media LLC},
  volume    = {17},
  number    = {1},
  pages     = {104},
  month     = jan,
  year      = {2025},
  doi       = {10.1007/s40820-024-01611-1},
  url       = {https://doi.org/10.1007/s40820-024-01611-1}
}

@article{Maiti2021-ps,
  title     = {Strain-induced spatially resolved charge transport in {2H-MoTe$_{2}$}},
  author    = {Maiti, Rishi and Saadi, Md Abid Shahriar Rahman and Amin, Rubab and Ozcelik, Veli Ongun and Uluutku, Berkin and Patil, Chandraman and Suer, Can and Solares, Santiago and Sorger, Volker J.},
  journal   = {ACS Appl. Electron. Mater.},
  publisher = {American Chemical Society (ACS)},
  volume    = {3},
  number    = {9},
  pages     = {3781--3788},
  month     = sep,
  year      = {2021},
  doi       = {10.1021/acsaelm.1c00281},
  url       = {https://doi.org/10.1021/acsaelm.1c00281}
}

@article{PhysRevLett.64.2074,
  title = {Magnetization of mesoscopic copper rings: Evidence for persistent currents},
  author = {L{\'e}vy, L. P. and Dolan, G. and Dunsmuir, J. and Bouchiat, H.},
  journal = {Phys. Rev. Lett.},
  volume = {64},
  issue = {17},
  pages = {2074--2077},
  numpages = {0},
  year = {1990},
  month = {Apr},
  publisher = {American Physical Society},
  doi = {10.1103/PhysRevLett.64.2074},
  url = {https://link.aps.org/doi/10.1103/PhysRevLett.64.2074}
}

@article{loss1998,
  author    = {Loss, Daniel and DiVincenzo, David P.},
  title     = {Quantum computation with quantum dots},
  journal   = {Phys. Rev. A},
  volume    = {57},
  number    = {1},
  pages     = {120--126},
  year      = {1998},
  month     = jan,
  publisher = {American Physical Society},
  doi       = {10.1103/PhysRevA.57.120},
  url       = {https://doi.org/10.1103/PhysRevA.57.120}
}

@article{miftah,
  title = {Effective $K$-valley Hamiltonian for transition metal dichalcogenide bilayers under pressure and application to twisted bilayers with pressure-induced topological phase transitions},
  author = {Anfa, Miftah Hadi Syahputra and Elatresh, Sabri and Bahlouli, Hocine and Vogl, Michael},
  journal = {Phys. Rev. B},
  volume = {111},
  issue = {24},
  pages = {245434},
  numpages = {15},
  year = {2025},
  month = {Jun},
  publisher = {American Physical Society},
  doi = {10.1103/5k9m-mfbz},
  url = {https://link.aps.org/doi/10.1103/5k9m-mfbz}
}

@article{AharonovBohm1959,
  author  = {Aharonov, Yakir and Bohm, David},
  title   = {Significance of Electromagnetic Potentials
             in the Quantum Theory},
  journal = {Physical Review},
  volume  = {115},
  number  = {3},
  pages   = {485--491},
  year    = {1959},
  doi     = {10.1103/PhysRev.115.485}
}

@article{Landauer1957,
  author  = {Landauer, Rolf},
  title   = {Spatial Variation of Currents and Fields
             Due to Localized Scatterers in Metallic
             Conduction},
  journal = {IBM Journal of Research and Development},
  volume  = {1},
  number  = {3},
  pages   = {223--231},
  year    = {1957},
  doi     = {10.1147/rd.13.0223}
}

@article{Buttiker1986,
  author  = {B{\"u}ttiker, Markus},
  title   = {Four-Terminal Phase-Coherent Conductance},
  journal = {Physical Review Letters},
  volume  = {57},
  number  = {14},
  pages   = {1761--1764},
  year    = {1986},
  doi     = {10.1103/PhysRevLett.57.1761}
}

@article{Groth2014Kwant,
  author  = {Groth, Christoph W. and Wimmer, Michael
             and Akhmerov, Anton R. and Waintal, Xavier},
  title   = {Kwant: A Software Package for Quantum Transport},
  journal = {New Journal of Physics},
  volume  = {16},
  pages   = {063065},
  year    = {2014},
  doi     = {10.1088/1367-2630/16/6/063065}
}

@article{Nikolic2006,
  author    = {Nikoli{\'{c}}, Branislav K. and Z{\^{a}}rbo, Liviu P. and Souma, Satofumi},
  title     = {Imaging mesoscopic spin {H}all flow: Spatial distribution of local spin currents and spin densities in and out of multiterminal spin-orbit coupled semiconductor nanostructures},
  journal   = {Phys. Rev. B},
  volume    = {73},
  issue     = {7},
  pages     = {075303},
  numpages  = {14},
  year      = {2006},
  month     = {Feb},
  publisher = {American Physical Society},
  doi       = {10.1103/PhysRevB.73.075303},
  url       = {https://link.aps.org/doi/10.1103/PhysRevB.73.075303}
}
\end{document}